\documentclass[prodmode,acmcsur]{acmsmall} 
\usepackage[bookmarks=false]{hyperref}

\newcommand{\ClusterHead}[1]{CH#1}
\newcommand{\BaseStation}[1]{GW#1}

\newcommand{\TableContinuous}[1]{\textbf{C}}
\newcommand{\TableQuery}[1]{\textbf{Q}}
\newcommand{\TableEvent}[1]{\textbf{E}}
\newcommand{\TableObject}[1]{\textbf{O}}

\usepackage{amsmath}
\usepackage{mathtools}
\usepackage{dsfont}
\usepackage{paralist}
\usepackage[table]{xcolor}
\usepackage{footnote}
\usepackage{array}
\usepackage{ragged2e}
\usepackage{bm}
\usepackage{textcomp}

\usepackage{rotating}

\usepackage{subcaption}

\usepackage{multirow}

\usepackage{pdflscape}
\usepackage{afterpage}

\usepackage{tikz}

\usepackage{float}
\restylefloat{table}

\usepackage{ifpdf}
  \DeclareGraphicsExtensions{.pdf}

\input{img-macros.aux}

\usepackage{setspace}

\usepackage{comment}
\usepackage{ifthen}

\doi{0000001.0000001}

\issn{1234-56789}

\begin{document}
%
%
%
%

\markboth{G. M. Dias et al.}{A Survey about Prediction-Based Data Reduction in 
Wireless Sensor Networks}

\title{A Survey about Prediction-based Data Reduction in 
Wireless Sensor Networks}
\author{Gabriel~Martins~Dias
\affil{Pompeu Fabra University}
Boris~Bellalta
\affil{Pompeu Fabra University}
Simon~Oechsner
\affil{Pompeu Fabra University}
}

\begin{abstract}

One of the main characteristics of Wireless Sensor Networks (WSNs) is the 
constrained energy resources of their wireless sensor nodes.
Although this issue has been addressed in several works and got a lot of 
attention within the years, the most recent advances pointed out that the 
energy harvesting and wireless charging techniques may offer means to
overcome such a limitation.
Consequently, an issue that had been put in second place, now emerges: the 
low availability of spectrum resources.
Because of it, the incorporation of the WSNs into the Internet of Things and 
the exponential growth of the latter may be hindered if no control over the 
data generation is taken.
Alternatively, part of the sensed data can be predicted without triggering 
transmissions and congesting the wireless medium.
In this work, we analyze and categorize existing prediction-based data 
reduction mechanisms that have been designed for WSNs. 
Our main contribution is a systematic procedure for selecting a scheme to 
make predictions in WSNs, based on WSNs' constraints, characteristics of 
prediction methods and monitored data. 
Finally, we conclude the paper with a discussion about future challenges 
and open research directions in the use of prediction methods to support the 
WSNs' growth.
\end{abstract}



\acmformat{Gabriel~Martins~Dias, Boris~Bellalta,~and~Simon~Oechsner, 2016. A 
Survey about Prediction-based Data Reduction in Wireless Sensor Networks.}

\begin{bottomstuff}
This work has been partially supported by the Spanish Government under project 
TEC2012-32354 (Plan Nacional I+D),  and by the Catalan Government 
(SGR-2014-1173).
\end{bottomstuff}

\maketitle

\section{Introduction}

%
%

Wireless sensor nodes (sensor nodes, for brevity) are small computer devices 
with low production costs, equipped with a radio antenna and sensors  that are 
capable of sensing one or more environmental parameters~\cite{Akyildiz2002}.
Thanks to their portable size, sensor nodes are often densely deployed in areas 
that may not be humanly accessible.
Hence, one of the biggest challenges of working with battery-equipped sensor 
nodes has been their limited energy availability, which is compounded by the 
fact that radio transmissions are the operations that consume the most energy 
and that Wireless Sensor Networks (WSNs) are mainly data-oriented networks, 
i.e., their most valuable asset is the data that sensor nodes can produce.

Nonetheless, a recent survey presented in \cite{Rault2014} listed several 
efforts to manage the energy consumption of WSNs at different levels.
In this survey, among several applications, routing and energy-saving 
techniques, there are promising advances in the energy supply methods for 
sensor nodes, including energy harvesting~(\cite{Sudevalayam2011}) and wireless 
power transfer~(\cite{Xie2013}).
According to \cite{Rault2014}, wireless power charging facilitates the design 
of scalable methods to refill network elements' batteries, allowing the 
sensor nodes' energy constraint to be overcome.
%
%
%
%
Meanwhile, the medium access is one of the key challenges in the next 
generations of wireless networks due to the increasing number of wireless 
devices and different traffic profiles~(\cite{Bellalta2015}).
Therefore, we foresee the urgency of reducing the number of transmissions to 
support the growth in the number of wireless devices, and the use of 
predictions is a promising alternative, also because they have potential as 
energy-saving mechanisms.
For the presented reasons, we survey existing approaches that use predictions to 
reduce the number of transmissions in WSNs.

Indeed, many prediction-based data reduction techniques have been designed for 
minimizing their radio energy consumption, without concerning about the medium 
access limitations.
For example, the survey presented in \cite{Anastasi2009} described several 
methods and architectures used to reduce the energy consumption and extend the 
WSNs' lifetime.
There, works are labeled according to their main characteristics, including 
a \emph{data-driven} category that encompasses a specific set of works 
focused on \emph{data prediction}.
As they focus on WSNs' energy consumption, some categories are 
overlapping in terms of how predictions are adopted and computed in WSNs.
Furthermore, as the authors highlighted, by the time of the publication, 
forecasting methods had not been fully explored in WSNs and only a very 
limited number of algorithms had been adopted, because high-complexity 
methods were thought to be unsuitable for sensor nodes.
Later, such an assumption started to be challenged, and real deployments 
incorporated advanced forecasting methods~(\cite{Aderohunmu2013}) and 
other artificial intelligence tools~(\cite{Askari2011}).
More recently, in~\cite{Rault2014}, several techniques to reduce the energy 
consumption in WSNs were compared according to the requirements of WSN 
applications.
Once again, works have been analyzed from the energy consumption perspective 
and, besides including several approaches that do not use predictions, the 
authors neither focus on how the predictions are adopted in WSNs nor which 
prediction methods are used to reduce the data.

Data reduction methods encompass different techniques that may lower the 
number of transmissions, but not all of them involve predictions.
For example, the mechanism presented in~\cite{Deligiannakis2011} prioritizes 
routes through sensor nodes that are collecting data at a certain time and 
paths that can aggregate more information thanks to the data similarity.
Hence, WSNs routing topology is optimized in favor to reduce their energy 
consumption and number of transmissions.
Alternatively, \cite{Intanagonwiwat2001} shows that a simple aggregation scheme 
that joins the data from packets and suppresses their headers' information
can efficiently reduce the number of transmissions in the WSNs.
The survey presented in~\cite{Luo2007} contains other examples of intelligent 
routing schemes, \cite{Fasolo2007} lists other data aggregation methods, and 
\cite{Srisooksai2012} is a survey of mechanisms for data compression.
All of these application types can reduce the number of transmissions, but 
prediction-based data reduction methods are not restricted to only one 
of them, as we will exemplify in Section~\ref{sec:sps}.

\afterpage{
\begin{landscape}
\begin{figure}[t]
        \centering
	\includegraphics[height=0.5\textwidth]{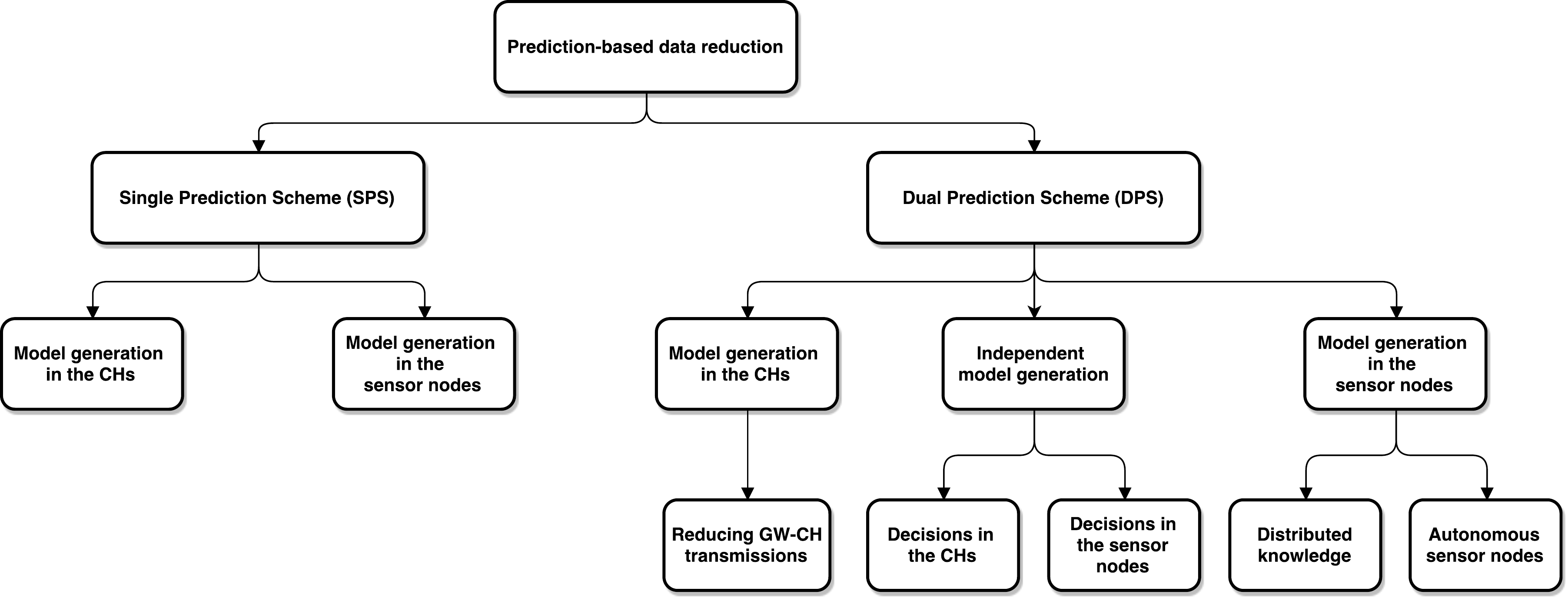}
        \caption{Taxonomy of the architectures that use predictions for data 
reduction.}
        \label{fig:taxonomy}
\end{figure}
\end{landscape}
}

Within the years and hardware evolution, some works started shifting the 
paradigm of avoiding complex algorithms in WSNs.
The works surveyed in~\cite{Mahmood2013} take into account the WSNs' 
constraints to adopt data mining techniques aiming to find patterns in the 
sensor data to improve its collection and delivery.
For instance, if nodes that measure similar information integrate the same 
cluster, redundant data can be efficiently suppressed, increasing the overall 
data delivery and improving the WSN's energy efficiency~(\cite{Guo2009}).
Even though some techniques adopted in data mining also include predictions, 
they are mainly applied to extract relevant information hidden in the sensed 
data, and many works included in the survey do not address the reduction in the 
number of transmissions.
Similarly,~\cite{Alsheikh2014} discussed the adoption of machine learning 
techniques at different layers, such as routing, medium access control and event 
detection.
In our work, a machine learning technique called Artificial Neural Networks 
(ANNs) will be also presented.
However, we also consider many statistical and probabilistic methods, such as 
the traditional time series methods (autoregressive and moving average methods), 
that are not considered machine learning.
While machine learning techniques rely on their ability of learning and 
evolving their predictions in response to environment changes, traditional time 
series methods rely on the statistics of the studied data to make predictions.

For the reasons explained at the beginning, we limit the scope of this survey 
to mechanisms that use predictions as a means to reduce the number of 
transmissions in a WSN.
Moreover, in our work, we focus not only on presenting the current solutions 
for WSNs, but also on introducing existing prediction techniques, featuring 
methods that are currently being used in data reduction solutions in WSNs.
To the best of our knowledge, this is the first time that this approach has 
been taken.



The rest of this paper is organized as follows, in 
Section~\ref{sec:wsn-environments}, we explain the terms and jargons that will 
be used to characterize WSNs and predictions in the rest of the work.
The following sections list the works that use predictions to reduce the number 
of transmissions, according to the structure shown in 
Figure~\ref{fig:taxonomy}: in Section~\ref{sec:sps}, we introduce and explain 
the Single Prediction Schemes (SPSs), featuring where the predictions are 
computed; and in Section~\ref{sec:dps} we detail the Dual Prediction Schemes 
(DPSs).
Section~\ref{sec:prediction-methods} shows the current state-of-the art from 
the perspective of the data predictions, i.e., which are the methods used to 
make predictions in WSNs, as well as the advantages and disadvantages of each 
one.
Later, we provide a discussion with the main questions and challenges observed 
in the surveyed works in Section~\ref{sec:discussion}, before listing, in 
Section~\ref{sec:future-challenges}, the issues that are still open in this 
area and the guidelines for new works that intend to solve such problems and 
improve the state-of-the-art.
Finally, we draw the conclusions in Section~\ref{sec:conclusion}.




\section{Fundamentals of WSNs and data prediction}
\label{sec:wsn-environments}

In order to make it clearer for the readers, we adopt a standard set of terms 
and describe all the surveyed works using the same nomenclature.
In this Section, we present the default representation considered for WSNs and 
explain the terms used to describe predictions, which may be crucial for 
understanding the rest of this work.

\subsection{WSN organization}

\begin{figure}[t]
        \centering
	\begin{subfigure}[t]{0.43\textwidth}
		\includegraphics[width=\textwidth]{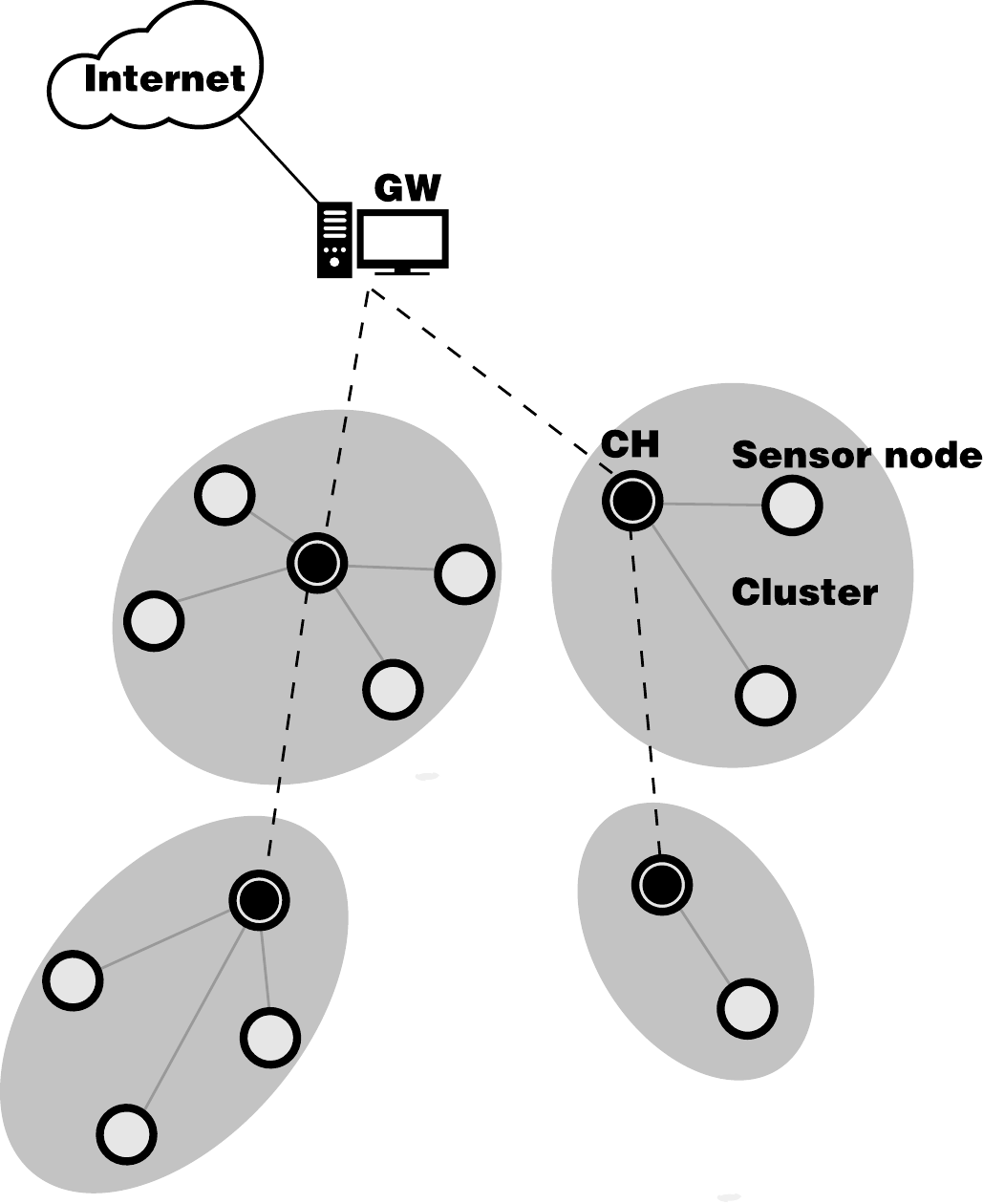}
		\caption{WSN with multiple clusters}
		\label{fig:scenario-wsn}
	\end{subfigure}
        \qquad
	\begin{subfigure}[t]{0.43\textwidth}
		\includegraphics[width=\textwidth]{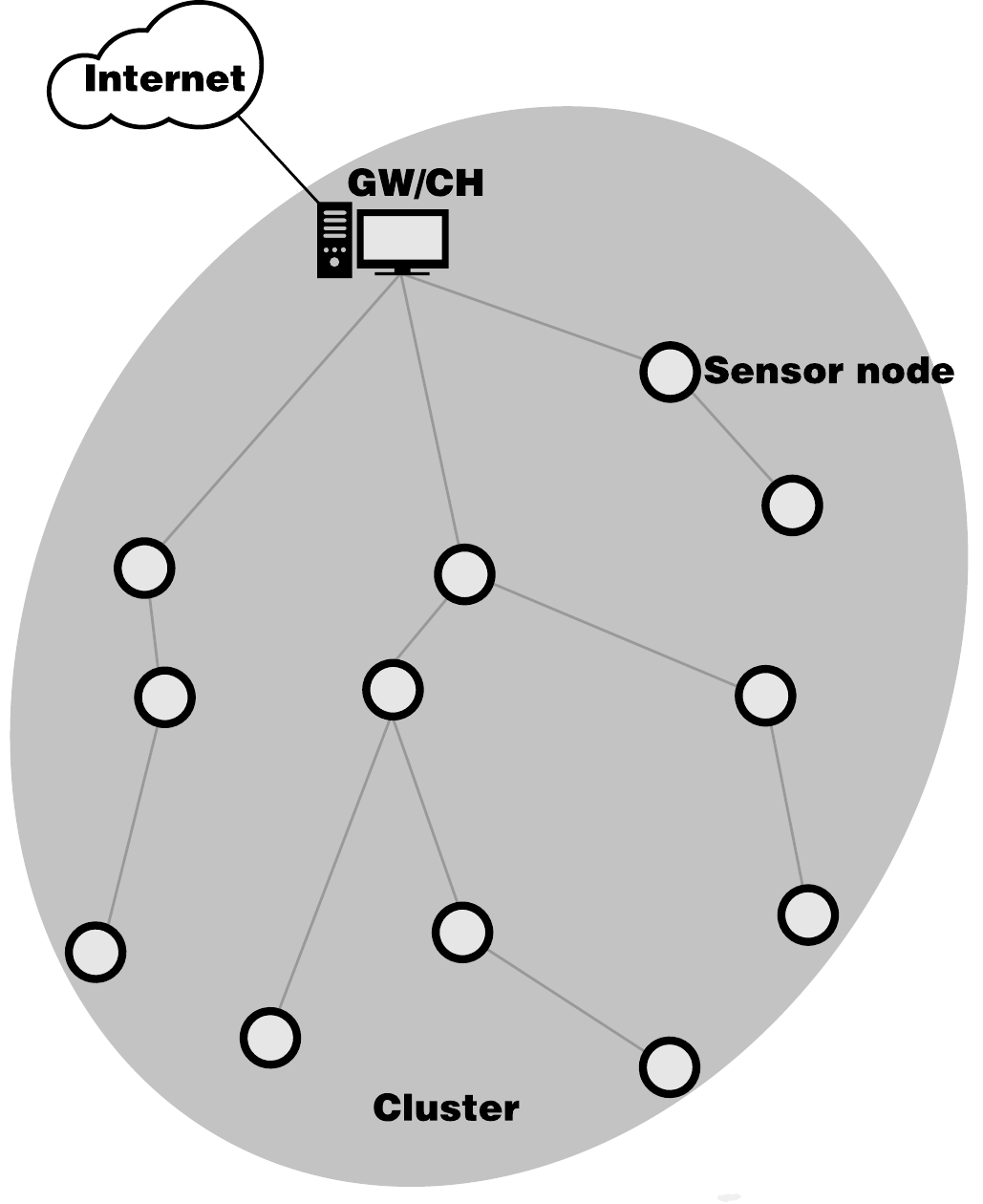}
		\caption{WSN without sensor nodes acting as Cluster Heads}
		\label{fig:scenario-wsn-2}
	\end{subfigure}
        \caption{Typical WSN scenarios and their roles.}
        \label{fig:wsn}
\end{figure}

A typical WSN is composed by dozens (occasionally hundreds) of ordinary 
sensor nodes connected to a central workstation that is responsible for 
providing the communication between the WSN owner and the sensor nodes.
This link is bidirectional: it can be used to inform the reported data to the 
WSN owner and to (re-)configure the sensor nodes' operation.
From now, we will refer to this central workstation as 
\emph{Gateway}~(\BaseStation{}).

In some cases, the excessive number of sensor nodes demands an internal 
WSN reorganization to avoid packet losses and reduce the number of packet 
collisions.
Such a internal organization, which facilitates the communication between 
sensor nodes and saves their batteries, is made by ``clustering'' sensor 
nodes according to their location or according to the correlation between their 
measurements (see~\cite{Abbasi2007} for further details). 
As shown in Figure~\ref{fig:scenario-wsn}, a WSN may be organized in one or 
several clusters. 

In a cluster, the communication between the sensor nodes and the \BaseStation{} 
is responsibility of the Cluster Head (\ClusterHead{}).
For example, \ClusterHead{s} must inform sensor nodes about decisions 
taken by the WSN owner and transmit the sensed data to the \BaseStation{}.
As the sensor nodes in a cluster are usually near to each other, it is easier 
for \ClusterHead{s} to keep the control of transmissions and reduce the sensor 
nodes' energy wasting.
Most of the works considered in this survey do not focus on ways to cluster 
sensor nodes, but they usually rely on existing methods to assume a 
cluster-based organization.

Alternatively to clustered WSNs, a flat organization may be adopted, as shown in 
Figure~\ref{fig:scenario-wsn-2}.
In such cases, \BaseStation{s} can communicate directly with sensor nodes and 
therefore assume the responsibility for establishing the communication between 
sensor nodes and WSN owners.
As \ClusterHead{s} and \BaseStation{s} have the similar responsibility of 
gathering data from sensor nodes and push it forward, from now, we will refer 
to \BaseStation{s} as \ClusterHead{s} in order to facilitate for the reader.
In specific cases where the communication between \BaseStation{s} and 
\ClusterHead{s} are discussed, it will be clearly stated.

\subsection{Data prediction}

The term \emph{prediction} can either refer to the process of inferring missing 
values in a dataset based on statistics or empirical probability, or to the 
estimation of future values based on the historical data.
A \emph{prediction method} ($P$) is a function that produces predictions based 
on two input values: a set of observed values ($X$) and a set of parameters 
$(\theta)$.
A \emph{prediction model} ($p$) is an instance of a prediction method $P$, 
such that $p_{\theta}(X) = P(X, \theta)$, i.e., every prediction model is 
deterministic and its output depends only on the set of observed values.
The values of $\theta$ can be chosen based on the evaluation provided by a 
utility function that can measure predictions' accuracy, models' complexity and 
information loss.
In conclusion, it is possible to create different prediction models that use 
the same algorithm (i.e., the same prediction method).

A prediction method may require some information about the data which is going 
to be predicted, for example, the assumption that the values will be normally 
distributed.
In some cases, this knowledge is already owned by the user before the 
deployment of the WSN and can be applied to statistical methods, such as linear 
regressions (see~\cite{Kong} for further details).
The positive aspect of statistical methods is that it is possible to estimate 
the yield of the system beforehand. 
For example, based on the assumption about the data normality, the probability 
of making accurate predictions can be used to calculate if the gains that 
the system can achieve will be worth the investment to be done.

On the other hand, machine learning techniques require fewer assumptions about 
the data in exchange for a period to adjust their parameters and adapt to the 
data that is being monitored, as described in~\cite{Haykin1999}. 
As a drawback, it is not possible to estimate how the system will perform in 
the real world before it is actually running. 
Usually, these techniques are tested through real data and adapt to occasional 
changes, but no guarantees about the predictions' accuracy can be given.

\section{Single Prediction Schemes}
\label{sec:sps}

In the Single Prediction Schemes (SPSs), predictions are made in a single point 
of the network, which can be either close to the origin of the data (in 
sensor nodes) or close to the data collection point (in \ClusterHead{s}). 
%
%
%
For instance, \ClusterHead{s} can predict the data measured by sensor nodes and 
autonomously decide when to pull more measurements based on the 
reliability of the predictions.
Alternatively, sensor nodes can predict changes in their surroundings to avoid 
unnecessary measurements and--consequently--their transmissions.
The latter option is especially beneficial if a sensor node spends more energy 
to sample the environment than to compute a set of machine instructions that 
will predict the future measurements.

The main advantage of SPSs is that each device can decide by itself whether to 
adopt predictions or not, and there is no overhead to communicate about their 
decisions or synchronize with their neighbors.
As a drawback, there is an eventual reduction in the quality of the information 
provided by \ClusterHead{s}, given that WSNs resign part of the data generated 
by their sensors.
In this Section, we categorize, according to the place where the predictions 
are made, existing works that adopt SPSs to reduce the number of transmissions.


\subsection{Model generation in the \ClusterHead{s}}

As \ClusterHead{s} usually have higher computational power and energy 
availability, they can locally generate prediction models and take important 
decisions about the WSNs' operation without compromising the quality of the 
information provided by the measurements.
On the other side, a conservative strategy is adopted in sensor nodes, 
which become merely responsible for their primary tasks, i.e., measuring 
environmental parameters and transmitting the raw data collected by their 
sensors.




Especially in environmental monitoring WSNs, measurements made by closely 
positioned sensor nodes have a spatio-temporal correlation, which can be used 
to generate probabilistic models, approximate the data to well-known 
distributions and associate confidence levels to predictions.
Hence, the number of transmissions can be reduced if \ClusterHead{s}
predict measurements and locally check whether the user-imposed quality 
constraints are matched or not.
Because of the autonomy given to the \ClusterHead{s}, this scheme have 
been used in several application types, such as adaptive sampling, clustering 
and data compression.

\paragraph{Adaptive sampling}

Generating predictions models in \ClusterHead{s} can be an efficient method to 
answer queries without fetching the data directly from the sensor nodes, as 
shown in~\cite{Cheng2003}.
User queries contain, besides the data that should be returned,
the error tolerated by the user. 
Therefore, \ClusterHead{s} can answer that the actual current measurements are 
inside a range of values if their confidence is high enough to satisfy an 
user-tolerated error.
To achieve that, \ClusterHead{s} must be able to compute prediction models
based on the statistics of the historical data--considering the uncertainty 
about the current values--and autonomously decide whether to pull more 
measurements or not.
As an alternative, \ClusterHead{s} can use inferential statistics to decide 
which sensor nodes have to be sampled, based on their odds of providing valuable 
information to the user.
%

In~\cite{Deshpande2004}, the mechanism called BBQ adopted linear regressions to 
exploit the correlation between different types of data that the sensor nodes 
may be able to measure, for example, their own voltage and the local 
temperature.
Simulations using real data show that the mechanism can reduce the number of 
transmissions, save energy and keep a high confidence level ($95\%$) about the 
information retrieved.
Moreover, its was possible to keep a low number of mistakes in a scenario 
with little human intervention, i.e., where the environment is influenced by 
fewer external factors.

More recently, Principal Component Analysis (PCA) was used to analyze the 
historical data and select only the sensor nodes that measured most of the 
variance observed in the environment~(\cite{Malik2011}).
The latter technique reduced the workload of the sensor nodes and prolonged 
twice the WSN lifetime, according the results obtained from experiments in real 
testbeds.

\paragraph{Topology control}
The works done in~\cite{Emekci2004,Yann-Ael2005} exploit the spatio-temporal 
correlation between the sensor nodes measurements to build sets of nodes that 
can provide ``trustful'' measurements and should be regularly sampled.
In practice, only a subset of sensor nodes is activated during a time 
interval and all the others have their radios and sensors turned off to reduce 
the number of transmissions, save energy and extend the WSN lifetime.
In order to fairly extend the WSN lifetime, every sensor node must be queried 
at least once during a cycle, and the number of times that they are activated  
in a cycle depends on the remaining energy on each one's battery.
On the other hand, every subset of sensor nodes provides the values used to 
predict the measurements of the whole WSN.
The predicted values, on average, should differ by less than a user-defined 
threshold from values obtained when using measurements from all the sensor 
nodes.
%
The Binocular framework (presented in~\cite{Emekci2004}) defines before the 
WSN deployment the subsets of sensor nodes that must be active at a time. 
During the so-called \emph{data processing} phase, \ClusterHead{s} receive 
measurements from sensor nodes and calculate linear transformations to make the 
predictions based on those sensor nodes that will remain active.
On the other hand, in~\cite{Yann-Ael2005}, measurements are assumed to 
follow normal distributions.
Simulations using real data show that their approach can be used to extend 
the WSNs' lifetime when the requirements about the accuracy are not very 
strict, 
namely, when the temperature can be wrong by $\pm0.5^{o}$C with a confidence 
level of $0.95$.
Furthermore, WSNs must be dense enough so that some sensor nodes can be
switched off and their measurements inferred using their neighbors' 
measurements.


\paragraph{Clustering}

The algorithm presented in~\cite{Tulone2006} introduced a method to build 
clusters and aggregate sensed data based on their similarity.
In short, nodes are accepted as part of a cluster if their measurements are 
similar to their neighbors' measurements, which reduces their divergences 
and the deviation from the average values, and makes the data easier to 
compress, for example.
Simulations using data collected by real sensor nodes showed that it is 
possible to reduce the number of transmissions done in the network without 
injecting significant errors to the reported data. 
%
%
The main drawback of this mechanism is that, as the sensor nodes' roles 
in the clusters rely on data analysis, it is not possible to assign the role of 
\ClusterHead{} based on the availability of resources, such as higher energy 
availability or higher computational power to perform advanced instructions.
This limitation impacts the solution presented in~\cite{Yin2015}, which uses 
the PCA to reduce the number of dimensions of the data and transmit less data 
from \ClusterHead{s} to \BaseStation{s}.
That is, \ClusterHead{s} may not have the computational power required to run 
PCA, because this method has high complexity and relies on some advanced 
instructions that cannot be performed in the simplest wireless sensor nodes, 
such as the multiplication of large matrices.


\subsection{Model generation in the sensor nodes}

If the prediction models are generated in the sensor nodes and not shared with 
the \ClusterHead{s}, their computing tasks may go far beyond the simple data 
reporting. 
For example, without the \ClusterHead{s}' intervention, they can decide if a 
measurement has to be made (or transmitted) based on the quality of the 
information that it can provide.

Given that the sensor nodes' computing power can be constrained, decisions about 
predictions may be supported by their neighbors' data, i.e., they can be 
distributed.
For instance, instead of transmitting every measurement to the \ClusterHead{}, a 
sensor node may locally decide to not transmit after observing that its 
neighbors measurements are sufficient to accurately monitor its region.
WSNs used for event detection and object tracking may predict in the sensor 
nodes to avoid burst transmissions that could provoke packet losses and delay 
the delivery of important messages to the \ClusterHead{s}.


Furthermore, WSNs for object tracking are usually composed by more powerful and 
reliable sensors, such as cameras, microphones and radio-frequency 
identification~(\cite{Bhatti2009}), in which the sampling process generally 
consumes more energy than the traditional temperature and relative humidity 
monitoring devices~(\cite{Anastasi2009}). 
Hence, predictions can be used to adjust their sampling rate and avoid 
unnecessary measurements, as proposed in
the Prediction-based Energy Saving (PES) scheme~(\cite{StrategiesXu2004}).
The mechanism defined by the PES has three main components: 
\begin{inparaenum}[(i)] 
	\item a simple prediction model that can be computed by the sensor 
nodes and avoid unnecessary computation;
	\item a wake up mechanism that defines which nodes should be turned on 
after making a prediction about where the object is going to be in the next 
time interval; and
	\item a recovery mechanism, in order to turn all sensor nodes on 
whenever an object which is expected to be in the range of WSN cannot be found 
by the 
active ones.
\end{inparaenum}
The PES scheme aims to minimize the miss rate while tracking objects and 
maximize the energy savings in the WSN, which is achieved by reducing the sensor 
nodes' computing time and their number of transmissions.
Simulation results showed that the success of this scheme depends mostly on the 
number of objects that will be tracked at the same time. 
Once this information is known, the dynamics of their movements 
play a major role on the system's efficiency.
In~\cite{Samarah2011}, the PES scheme was simulated using a multi-dimensional 
regression analysis to predict the movements of the tracked objects.
The simulation results showed that it was possible to keep a low energy 
consumption level while maintaining the missing rate less than $20\%$.



\section{Dual Prediction Schemes}
\label{sec:dps}

\begin{figure}[t]
	\centering
	\includegraphics[width=0.9\textwidth]{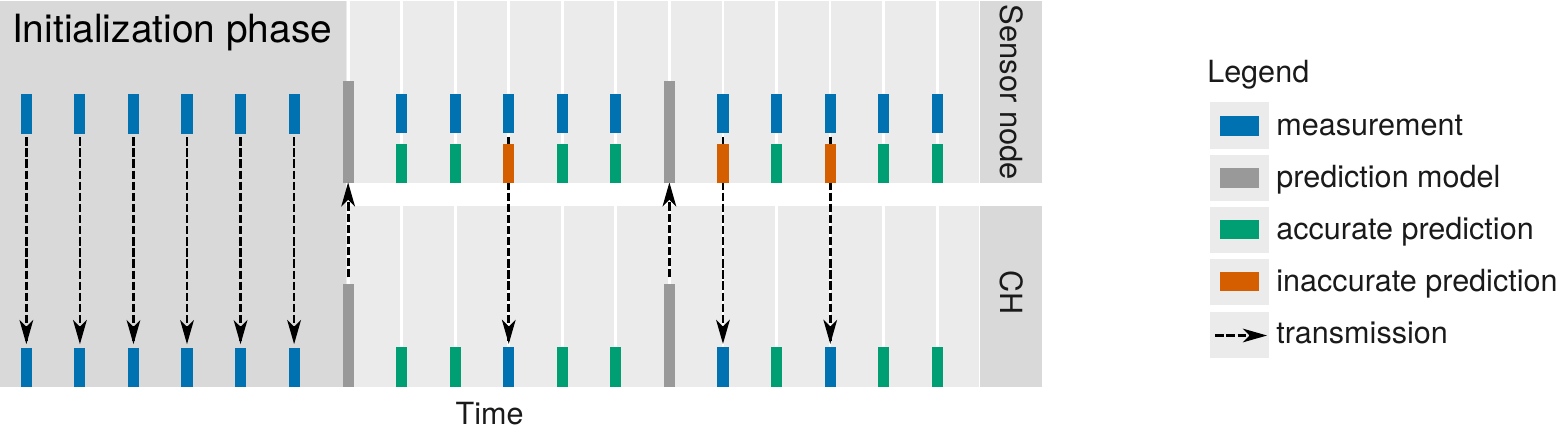}
	\caption{In a DPS, a measurement is transmitted only if its
	forecast is inaccurate. The \ClusterHead{s} may be responsible for 
transmitting new prediction models every time interval after the 
initialization phase.}
	\label{fig:timeline-dps-ch}
\end{figure}
\begin{figure}[t]
	\centering
	\includegraphics[width=0.9\textwidth]{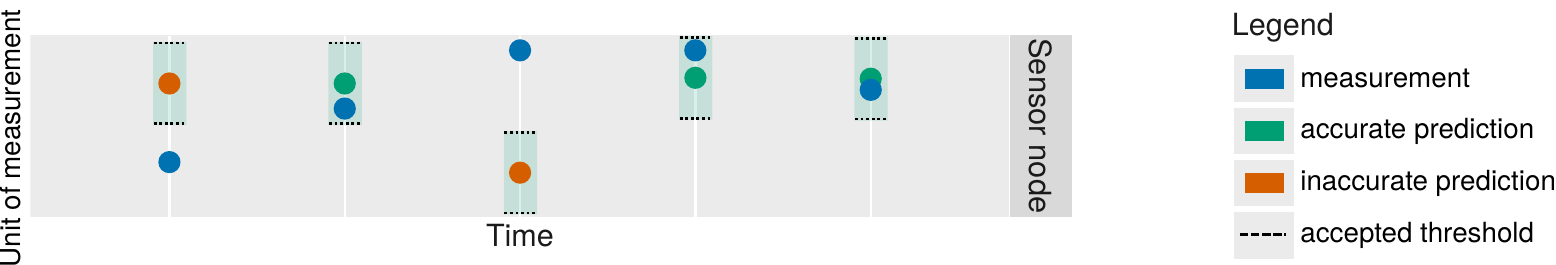}
	\caption{Measurements that fall inside the accepted threshold do not 
trigger any action}
	\label{fig:trigger-threshold}
\end{figure}

In Dual Prediction Schemes (DPSs), the predictions are simultaneously made in 
\ClusterHead{s} and sensor nodes.
The general idea behind such mechanisms is that sensor nodes are able to produce 
the same ``a priori'' knowledge as \ClusterHead{s} are, but sensor nodes 
can locally check the predictions' accuracy and avoid unnecessary transmissions.
As shown in Figure~\ref{fig:timeline-dps-ch}, the same prediction model is 
shared between the sensor node and its respective \ClusterHead{}.
Then, every time the sensor node measures a value that falls outside an 
acceptance threshold defined for the predictions (as represented by the first 
and the third measurements in Figure~\ref{fig:trigger-threshold}), it must 
transmit the real value to its \ClusterHead{}, which substitutes the local 
predictions by the correct value.
Hence, sensor nodes can consume less energy resources and avoid unnecessary 
transmissions, because measurements will be transmitted to the \ClusterHead{s} 
only when the predictions are not sufficiently accurate.

Prediction models can be generated in \ClusterHead{s} and further shared with 
sensor nodes, or vice-versa (for instance, the work presented 
in~\cite{Lazaridis2003} allows both approaches).
Once in a while, a new prediction model may be generated if the current one is 
not predicting as accurate as expected.
As DPSs aim to reduce the number of transmissions without compromising the 
quality of information generated by the WSN, they target the trade-off 
between the number of transmissions and the quality of measurements provided by 
the system.
In this case, DPSs' efficiency depends not only on predictions' accuracy, but 
also on the number of transmissions required to distribute new prediction 
models and on the channel's reliability.
For instance, a high bit error rate may result on absence of updates arriving at 
the \ClusterHead{s}, which is usually treated as a signal of predictions' high 
accuracy.
Hence, to avoid that, packets will have to be often retransmitted, congesting 
the medium, consuming extra energy and diminishing the theoretical gains.
Therefore, to decide for a new prediction model, it is necessary to observe 
what is the most proper prediction method, given the current environmental 
conditions, and if making predictions (instead of transmitting all measurements) 
will reduce the number of transmissions in the WSN.
These observations may be made either in \ClusterHead{s} or in sensor nodes, 
independently of their responsibility to generate new prediction models, in 
order to keep the high accuracy and the scheme's yield.

Alternatively, a sensor node and its \ClusterHead{} may generate the same 
prediction model, independently and at the same time, without the necessity of 
generating extra transmissions in the WSN.
This requires a previous knowledge about the environment and the data that is 
going to be measured, in order to program and configure sensor nodes to decide 
for the same methods as \ClusterHead{s} will adopt in runtime.
Furthermore, in this case, the predictions' accuracy may be restricted by the 
sensor nodes' computing capacity, because sensor nodes may not be able to adopt 
prediction methods that require more memory, storage or processing power.

%
%

\subsection{Model generation in \ClusterHead{s}}

Generating the prediction models in \ClusterHead{s} exploits the asymmetric 
computational power availability in WSNs: \ClusterHead{s} usually have 
cheaper energy sources and more resources (such as memory and processing power) 
than ordinary sensor nodes that are mainly used for measuring and reporting 
environmental data.
As presented in~\cite{Goel2001,Liu2005}, at the beginning, sensor nodes transmit 
the current measurements to their \ClusterHead{s}. 
Based on received values, \ClusterHead{s} are able to locally generate new 
prediction models for each sensor node.
Thus, \ClusterHead{s} are responsible for periodically updating and transmitting 
new prediction model parameters and error acceptance levels to 
their sensor nodes, as shown in Figure~\ref{fig:timeline-dps-ch}.
In~\cite{Li2009}, \ClusterHead{s} are also responsible for ensuring that 
sensor nodes have not stopped working, which may represent, in real deployments, 
a significant increase in the number of transmissions, besides overloading the 
network in case of dynamic scenarios or narrow tolerance for errors.

%

\subsubsection{Decisions in \ClusterHead{s}}

The Dual Kalman Filter (DKF--presented in~\cite{Jain2004}) uses spatial 
correlation between measurements from different sensor nodes.
Then, the algorithm assesses to the \ClusterHead{s} the ability (and the 
responsibility) of computing several prediction models for each sensor node and 
choose the best one according to the measurements received.
Predictions are made using a modified (distributed) version of the Kalman 
Filter that takes into account all information that \ClusterHead{s} may 
have, especially from other sensor nodes.
Similarly, the Efficient Data Gathering in Sensor Networks (EDGES--presented 
in~\cite{Min2010}) incorporates not only spatial, but also temporal correlations 
between received measurements.

\subsubsection{Decisions in sensor nodes}

The mechanism proposed in~\cite{Kho2009} gives to sensor nodes the 
ability of taking decisions locally using GP regression.
There, each sensor node needs to predict the information that is going to be 
sampled and adjusts its sampling schedule, according to the energy constraints, 
in order to maximize the information that it will collect during a particular 
time interval.
Simulation results over the data collected from real sensors showed that it is 
possible to maximize the quality of the information produced by the WSN and 
reach a high level of confidence by sampling as often as possible, with the 
constraints imposed on the limited available power in the sensor nodes.


\subsubsection{Reducing the transmissions between \ClusterHead{s} and 
\BaseStation{s}}

The Prediction-based monitoring (PREMON)~\cite{Goel2001} exploits 
spatio-temporal correlations between measurements from different sensor 
nodes to predict readings that would be done by some sensors, and to reduce the 
number of transmissions done by sensor nodes to their \ClusterHead{s} in a DPS. 
Furthermore, to reduce the number of transmissions between \ClusterHead{s} 
and \BaseStation{s}, \ClusterHead{s} may transmit only the prediction 
model and the updates to the \BaseStation{}, instead of the (aggregated) data 
that is usually transmitted.
Given that \BaseStation{s} and \ClusterHead{s} often have higher computational 
power availability, this kind of architecture supports more 
robust prediction methods and machine learning techniques.
For instance, \cite{Wu2016} used PCA to reduce the number of dimensions of 
the data transmitted from \ClusterHead{s} to \BaseStation{s}.
The main drawback is that the data in the \ClusterHead{s} is approximated, 
i.e., it may contain (small) errors introduced by the predictions used to 
reduce transmissions from sensor nodes, and these errors will propagate to 
the \BaseStation{s}.
To avoid risks of error propagation, it is possible to adopt a more 
conservative scheme, such as the mechanism called Ken (presented 
in~\cite{Chu2006}).
Ken differs from the previous methods, because the number of transmissions made 
by sensor nodes to \ClusterHead{s} is not reduced.
Instead, predictions are done in \ClusterHead{s} and in \BaseStation{s}, and 
values aggregated in \ClusterHead{s} are transmitted if predictions are 
inaccurate. 


\subsection{Independent model generation}

\begin{figure}[t]
	\centering
	\includegraphics[width=0.9\textwidth]{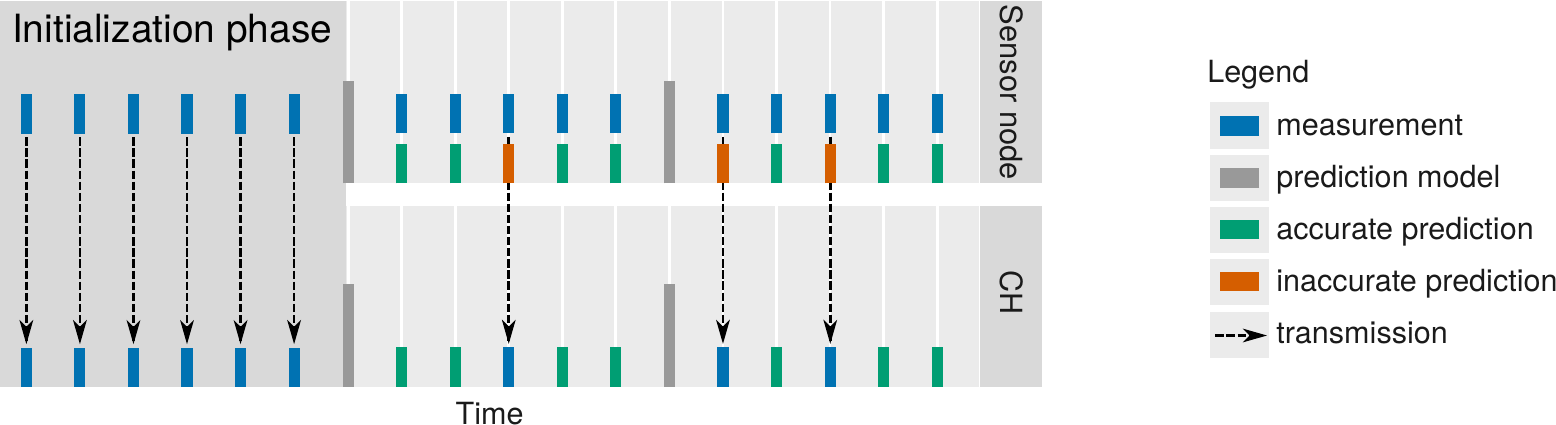}
	\caption{A variant of the DPS with independent model generation. It is 
not necessary any communication between the sensor node and the \ClusterHead{} 
to compute the same prediction model, because they are programmed to use the 
same data.}
	\label{fig:timeline-dps-independent}
\end{figure}

An independent model generation relies on an ``initialization 
phase'', i.e., a period during which sensor nodes report all the data that they 
have generated to \ClusterHead{s} (\cite{Santini2006}).
The initialization phase ensures that \ClusterHead{s} will have complete 
information about the environment before any prediction model is generated. 
After the initialization, \ClusterHead{s} are able to generate the same 
prediction models generated in their sensor nodes without making any extra 
transmission. 
At this moment, both start predicting the values, with the advantage that the 
sensor nodes are able to locally verify if a prediction is inaccurate and 
transmit the actual measurement, if needed.
Hence, sensor nodes may either regularly report the data to their 
\ClusterHead{s} due to the lack of accuracy in predictions, or not report 
any sensor reading at all, in case that the predictions are sufficiently 
accurate.
Figure~\ref{fig:timeline-dps-independent} illustrates the sensor nodes' and the 
\ClusterHead{}'s behaviors.

In~\cite{Santini2006,Wu2016}, the least mean squares method was used to predict 
future measurements, which was extended and improved in the simulations made 
in~\cite{Stojkoska2011}.
Meanwhile, \cite{Debono2008,Aderohunmu2013Impl,Aderohunmu2013} showed 
results of an implementation using real sensor nodes.
Especially, \cite{Aderohunmu2013Impl,Aderohunmu2013} compared the savings using 
several prediction methods: the constant method, weighted MAs, ARIMAs and the 
ES.
According to their results, the constant prediction method was the best 
trade-off between accuracy and energy consumption in sensor nodes.

\subsubsection{Decisions in \ClusterHead{s}}

According to the mechanism proposed in~\cite{Jiang2011}, \ClusterHead{s} can 
adapt the sensor nodes' operation according to the potential savings that
predictions may introduce.
To decide it, the authors use a formula to calculate whether it is worth to 
make predictions in sensor nodes or not, based on the relation between the 
predictions' accuracy, the correlation between measurements and the error 
tolerated by the user. 
According to the estimated gains, sensor nodes can be set to: 
\begin{inparaenum}[(i)] 
	\item go to sleep mode, without making any measurement; 
	\item make measurements and transmit every measurement done; or
	\item make measurements, transmit them to the \ClusterHead{} whenever 
the prediction differs by more than an accepted value, and update 
the prediction model parameters when necessary.
\end{inparaenum}

\subsubsection{Decisions in the sensor nodes}

In~\cite{Marbini2003,Ragoler2004,Jain2004a}, sensor nodes may have the ability 
to make further decisions based on the predictions' accuracy.
In~\cite{Ragoler2004}, sensor nodes can decide to aggregate the data 
received from their neighbors by suppressing measurements that are inside their 
confidence interval, instead of forwarding them to the \ClusterHead{s}.
To make such savings possible, \ClusterHead{s} must make the same 
predictions in order to answer to user queries locally.
In~\cite{Jain2004a}, sensor nodes may locally decide whether to 
adjust their sampling rate or not, based on the accuracy of their predictions. 
In theory, while local predictions are accurate, their sampling rate can be 
reduced and their energy can be saved by turning off their components for 
longer intervals. 
In this case, \ClusterHead{s} are responsible for controlling the bandwidth 
consumption to avoid peaks of packet transmissions and, consequently, 
collisions.

The works done in~\cite{Raza2012,Raza2015} propose a new naive algorithm for 
predicting. 
It is a linear approximation that uses recent measurements to calculate the 
slope of the measurements' trend. 
Predictions can be easily calculated by sensor nodes using interpolation.
According to their simulation results, the suppression ratio can reach high 
levels (up to $99\%$ of the application data) if the data has low variability.
In a real world test bed, the energy savings were significant (nearly $85\%$ of 
reduction), even though they did not reach the suppression ratio levels that 
the simulations had suggested.

\subsection{Model generation in sensor nodes}

\begin{figure}[t]
	\centering
	\includegraphics[width=0.9\textwidth]{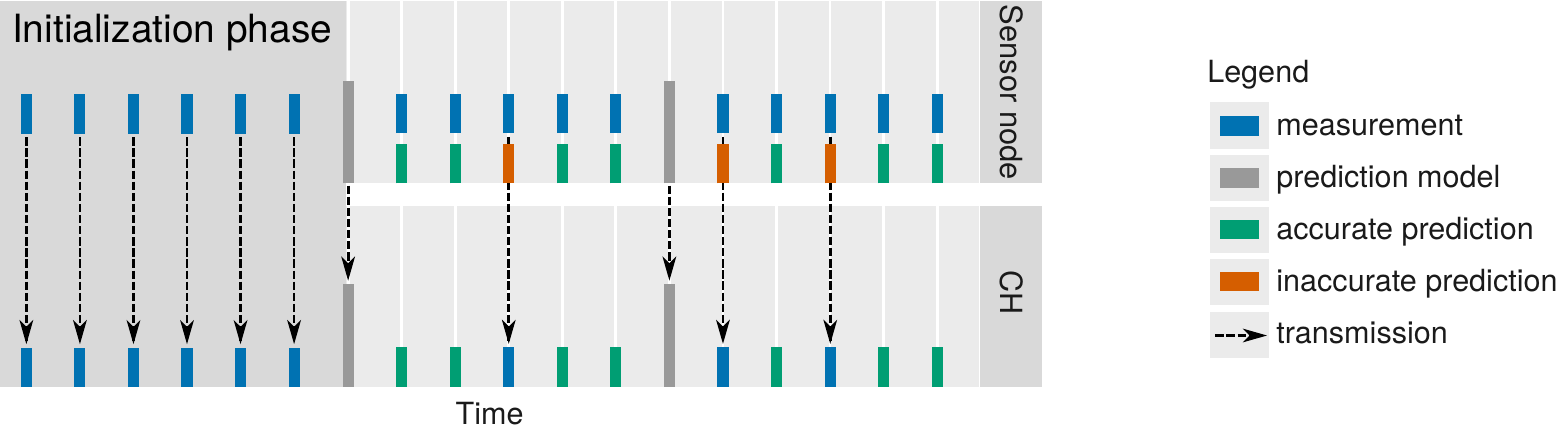}
	\caption{As sensor nodes can overhear their neighbors' data without 
overloading the network or congesting the medium, they may locally decide the 
best prediction method and later inform the computed model to 
the \ClusterHead{s}.}
	\label{fig:timeline-dps-sensor-node}
\end{figure}

The main drawback of generating prediction models independently was 
addressed in~\cite{LeBorgne2007}: approaches that rely on pre-defined 
prediction methods can lead to poor prediction performances if the model choice 
is not accurately done.
The authors decided to generate prediction models in sensor nodes (and not in 
\ClusterHead{s}, the alternative solution), as shown in 
Figure~\ref{fig:timeline-dps-sensor-node}.
As in the other DPSs, sensor nodes start transmitting all the measurements to 
their respective \ClusterHead{s}.
However, a new responsibility is assigned to sensor nodes: 
after collecting local measurements, they must fit a prediction model to the 
real data and communicate any occasional change to their \ClusterHead{s}.
Fitting a prediction model means finding the model that best summarizes the 
real measurements.
Hence, this mechanism requires much more computing power from the 
nodes than the other approaches (both to store more data and to choose the 
prediction models) and the savings depend on the predictions' accuracy, 
which may vary according to the sensed phenomenon and the data sampling rate.
Moreover, the choice of the prediction method is restricted by the memory and 
process power limitations of the sensor nodes.

\cite{LeBorgne2007} tested how better AR models can improve the WSN lifetime 
in comparison with constant prediction models.
The results of the simulations using real data from WSNs showed that the 
adaptive approach can reduce the number of data transmissions with neither 
exceeding the constrained memory nor the computational resources of common 
wireless sensor nodes.
With identical architectures, \cite{Li2013} chose the traditional ARIMA and 
\cite{McCorrie2015} the ES method to predict temperatures in the environment 
and in aircraft engines, respectively.
The hybrid model presented by \cite{Askari2011} improved the quality of the 
predictions giving to sensor nodes the autonomy to adopt an ANN when the 
predictions using ARIMA were inaccurate. 
In the worst case, if the predictions using the ANN model also fall outside the 
accepted threshold, sensor nodes are responsible for transmitting the real 
measurements and the new models' parameters to their \ClusterHead{s}.

Additionally to the predictions,~\cite{Lattanzi2013,Bogliolo2014} 
adopted a solution at hardware level (called wake-up receiver) to reduce the 
energy consumption during idle periods, improving the communication between 
sensor nodes.
Simulation results showed that the combination of both mechanisms could lead to 
larger gains than adopting each technique in isolation.
The main drawback is that the WSNs' topology cannot change when hardware 
solutions are adopted, because these approaches uses directional capsules 
and might stop working if the sensor nodes move. 

\subsubsection{Distributed knowledge}

Some works adopt the strategy of disseminating the sensor nodes' knowledge to 
their neighbors.
On the one hand, having more information about the surroundings gives to 
sensor nodes the ability to compute better prediction models before 
transmitting their parameters to the \ClusterHead{s}. 
Therefore, more accurate predictions reduce the number of long distance 
transmissions between sensor nodes and \ClusterHead{s} that usually 
represent a reasonable overhead in the overall number of 
transmissions~(\cite{Lee2003}).
On the other hand, extra information sharing may represent a waste of resources, 
because some sensor nodes will never be able to track an object (or a 
phenomenon) due to their distance or because the extra information is simply not 
sufficient to improve their predictions.

The dual prediction-based reporting mechanism (DPR) was firstly introduced in 
\cite{Lee2003} and further explored in~\cite{Xu2004}, using a prediction-based 
approach for performing energy efficient reporting in object tracking sensor 
networks. 
The data received from neighbors can be used by a sensor node to predict if 
an object will be in its range and activate its sensors only if necessary.
If its sensors are activated, a sensor node can verify whether its 
predictions about the object position were accurate and notify the 
\ClusterHead{} about occasional inaccuracies.
For the simulations, some prediction methods, such as the constant, the MA and 
the ES method, were adapted to the sensor nodes' limitations to reduce 
computational costs.
The simulation results showed that the algorithm is mainly affected by the 
reporting frequency, but the dynamics of the objects' movements are also 
important and reflect on the energy saved by these approaches.

The approach in~\cite{Guestrin2004} was designed for WSNs that monitor 
environmental parameters.
It relies on the correlation between measurements done by sensor nodes that are 
closely placed, which requires their exact localization. 
Firstly, the distributed mechanism creates a junction tree (explained 
in~\cite{Paskin2003}) layer on the top of the routing layer to disseminate the 
prediction method.
During the \emph{dissemination phase}, sensor nodes receive from \ClusterHead{s} 
some information according to their physical location, which 
may be used to calculate the prediction model parameters.
Thus, each sensor node is responsible for calculating part of a linear system 
using the available information at the time and broadcasting the (partial) 
results to its neighbors.
This is done until every node gets access to the complete information from 
all other nodes in the same cluster. 
Once such information is completely disseminated, the parameters of the 
prediction models are transmitted to the \ClusterHead{s}, which makes them able 
to predict new measurements and reduces the number of transmissions from 
sensor nodes.
Finally, whenever a sensor node detects that the difference between the 
predicted value and the measurement is greater than a fixed threshold, 
it updates the prediction model parameters.
Simulation results using static sensor nodes showed that this mechanism was 
able to reduce the number of transmissions and save energy, verifying also that 
it is highly scalable. 
The authors suggest that it can be extended to make WSNs support hundreds of 
nodes and run different applications, such as detections of outliers, data 
compression, and adaptive data modeling.
Moreover, even though~\cite{Guestrin2004} adopted the kernel regression to 
predict new measurements, they argued that the distributed mechanism may be 
extended to support other techniques.

\cite{Garrido-Castellano2015} implemented in real wireless sensor nodes the 
distributed kernel least squares regression method (explained 
in~\cite{Predd2009}).
There, each sensor could predict the temperature measurements of its neighbors 
based on their position and the local measurements.
Differently from the mechanism presented in~\cite{Guestrin2004}, the authors 
focused on the prediction algorithm and explained its utility in scenarios 
where sensor nodes must take local decisions based on complex and non-linearly 
distributed data.
Although the implementation has worked and the algorithm showed the same 
precision as in the simulation results, the authors listed several problems to 
deploy the mechanism in a real WSN. 
Among other problems encountered, they gave special attention to sensor nodes' 
small memory size and their restricted computing capabilities, which required 
a workaround to handle floating point operations.

The approach presented in~\cite{Carvalho2011Journal} also distributes tasks 
among sensor nodes. 
However, instead of requiring the exact position of each sensor node, they rely 
on their proximity to find correlated measurements.
Each node is responsible for calculating a linear regression based on its 
different measurement types.
For instance, a sensor node equipped with temperature and relative humidity 
sensors may use measurements of temperature to predict the relative humidity 
at a certain time.
After calculating the regression coefficients, each node broadcasts them to its 
neighbors.
Every neighbor that observes the same coefficients may inform the 
\ClusterHead{} that its measurements are similar.
\ClusterHead{s} receive coefficients and predict future measurements and, 
from this moment, a sensor node will only transmit the measurements if the 
predictions are inaccurate.
The simulation results showed an improvement in the energy consumption and a 
reduction in the number of transmissions, compared with the default operation.
However, the authors did not take into consideration the energy consumption to 
calculate the coefficients of the linear regression, nor the extra space 
used to store the measurements for the regression analysis.


\subsubsection{Autonomous sensor nodes}

Some authors propose that sensor nodes should have complete autonomy and 
decide by themselves, based on predictions, if they should pause on making 
measurements for a period.
\cite{Shen2008} developed a way to calculate whether sensor nodes should turn 
some of their components off or not, based on their total amount of energy and 
the time required to switch off their components. 
Moreover, each sensor node is able to calculate the possible amount of energy 
saved and to decide when to change its internal status.
That is, based on the predictions, a sensor node only changes its internal 
status if some energy will be saved.
For the predictions, they use Wavelet Neural Networks--an extension of ANNs.

\section{Comparison of the Prediction Methods Adopted in WSNs}
\label{sec:prediction-methods}

In the previous sections, we presented and discussed about the architectures 
and schemes that support the adoption of prediction methods in WSNs.
Moreover, we highlighted the degree of autonomy that \ClusterHead{s} and sensor 
nodes may have in some cases, which impacts the decision about the prediction 
method used.
For instance, a scheme that forces sensor nodes to work autonomously 
may not achieve the desired gains if a computationally intensive prediction 
algorithm 
is adopted.
On the other hand, a scheme that exploits the extended computational power of 
the \ClusterHead{s} would be sub-utilized if a simplistic (so-called 
\emph{naive}) method was adopted.

Therefore, before discussing the aspects that should be taken into consideration 
before adopting a prediction-based data reduction scheme in a WSN (in the 
next Sections), we present the prediction algorithms that have been used in 
WSNs.
At first, we split them into three classes: time series methods, 
regression methods and machine learning techniques.
For each algorithm, we highlight its potential to reduce the number of 
transmissions according to the data characteristics, before 
discussing other particularities that may impact its adoption in WSNs, such 
as time and space complexities.

\subsection{Time series methods} 

A time series is a sequence of data points, typically consisting of 
observations made over a time interval and ordered in time (\cite{Box2008}). 
Each observation is usually represented as $x_t$, where the observed value $x$ 
is indexed by the time $t$ in which it was made.
Thus, a \emph{time series prediction model} uses a time series as input to 
make predictions. 
These predictions are represented as a function of the past observations and 
their respective time, i.e., $x_{t} = f(x_{t-1}, \ldots, x_{t-k})$, where the 
function $f$ is a prediction model usually defined by parameters calculated 
using past observations. 
The algorithms used to find acceptable values for the parameters (i.e., those 
that may generate accurate predictions) may require some extra computation 
before making any prediction. 
Furthermore, since the environment may evolve and change, the parameters used 
to define a model may become obsolete after a while and hence the predictions' 
accuracy may decrease. 
Therefore, there is a computational cost to update the parameters of the chosen 
method and keep the predictions' accuracy. 

In the following, we explain the time series methods (also called 
\emph{forecasting methods}) used in WSN environments: 
the naive approaches, the Autoregressive (AR), the Moving Average (MA), the 
Exponential Smoothing (ES) and the Autoregressive Integrated Moving Average 
(ARIMA).
Table~\ref{table:time-series-methods} summarizes their characteristics and the 
reader can refer to \cite{Makridakis1998,Box2008,hyndman2014forecasting} 
for detailed information on forecasting methods.

\paragraph{Pros}

The main advantage of time series methods is their independence.
That is, it is not necessary neither external data nor an extended analysis to 
make accurate predictions. 
Moreover, each wireless sensor node may be able to predict short time intervals 
without depending on the support from neighbors' or \ClusterHead{s}' computing 
power, due to the low space-time complexity of such methods.


\paragraph{Cons}

Most of the time series methods adopted in WSNs assume that there is only 
one data type and neglect the presence of multiple sensors in a node, which is 
not uncommon and may be better exploited by other methods.
Furthermore, their accuracy is (usually) significantly lower when they are 
used for long-term predictions. 
That is, the data usually has similar values in the near future, but unobserved 
phenomena and sensor nodes' constraints may have a long-term impact, affect the 
predictions' accuracy and lead to fundamental errors, such as negative 
dimensions.

\input{table-methods.aux}

\subsubsection{Naive approaches}

Examples of naive approaches for predictions may vary between the 
\emph{average} of the past observations, the \emph{maximum} observed value or 
exactly the \emph{same} as the last observation made in time. 
Thanks to their simplicity, even though the predictions are not the most 
accurate they could be, naive approaches are usually compelling options for 
WSNs composed by sensor nodes with energy constraints or low computing power.


\paragraph{Pros}

There is no complex data processing in naive approaches, which makes them an 
affordable option in terms of computing time and required memory space.
Typically, naive approaches assume that future values can always be 
computed in constant time, given the historical data. 

\paragraph{Cons}

If the data has high variability, these methods are usually more inaccurate 
and imprecise than the most advanced methods.
In other words, if the tolerated error threshold is small, they will fail most 
of the time. 
Otherwise, they will fail less often, but their precision will be also reduced, 
because they will consider a wider range of values to assess predictions as 
accurate.

\subsubsection{Autoregressive (AR)}

Regressive models are used to represent the expected value of a variable given 
the values from a set of correlated variables.
In an autoregressive (AR) model, the mean and the variance of the values are 
constant, and the estimation is a linear combination of the values from the 
same variable. 
An AR model with order $p$ (referenced as $\text{AR}(p)$) is defined by a set of 
$p$ coefficients $\alpha_1,~\ldots,~\alpha_p$.
The parameter $p$ is usually set among $k$ possible values, using an 
information criterion measure (usually, $k \in [0,6]$).
Then, a value observed at time $t$ can be represented as

\begin{equation}
	x_{t} = c + \sum_{i=1}^{p}{\alpha_{i} x_{t-i}} + \varepsilon_t,
	\label{eq:ar}
\end{equation}
where the term $\varepsilon_t$ is a Gaussian white noise with variance 
$\sigma^2$, and $c$ is a constant such that, in case of having a 
stationary process with mean $\mu$,

\begin{equation}
c = \mu (1 - \sum_{i=1}^{p}{\alpha_{i}}).
\end{equation}

The value $\hat{x}_{t+1}$ can be predicted as $\hat{x}_{t+1} = c + 
\sum_{i=1}^{p}{\alpha_{i} x_{t+1-i}}$ using the same set of parameters 
$\alpha_{i}$.

\paragraph{Pros}

A trend observed in the (relatively) near past can be used to improve the 
predictions about the near future.
Moreover, the AR method is very efficient for short-term predictions for two 
reasons: 
\begin{inparaenum}[(i)]
 \item it is less sensitive than the naive predictions against the data 
noise; and
 \item its predictions follow trends observed in the most recent observations, 
which represents a potential to keep a high accuracy even if the data has high 
variance.
\end{inparaenum}

\paragraph{Cons}

Long-term predictions using AR tend to be inaccurate, due to uncertainty 
about the order of the model, its coefficients and unobserved errors.
Especially when the predicted period is longer than the order of the model,
predicted values are used for making new predictions, which propagates errors 
and affects the overall accuracy.

\subsubsection{Moving Average (MA)}

Similarly to the AR models, a MA model is defined by an order $q$ and is 
referenced as $\text{MA}(q)$.
Its order is defined as a window length $q$ that represents the number 
of past measurements that will be taken into account in the predictions. 
As in the AR method, the value of $q$ is usually set among $k$ possible values 
using an information criterion measure (usually, $k \in [0,6]$).

During the learning phase, which takes at least $q$ time intervals, the 
algorithm stores the measured data that will be used to calculate the model 
parameters and make the predictions.
In order to incorporate eventual changes in the future predictions, the MA 
method calculates the weighted average of the observations recently made.
The prediction of the value of $x$ at time $t$ is calculated using the 
following formula:

\begin{equation}
 x_{t} = \mu + \varepsilon_t + \sum_{i=1}^{q}{\theta_i \varepsilon_{t-i}},
\end{equation}
where $\mu$ is the average of the last $q$ values,
\begin{equation}
 \mu = \sum_{i=1}^{q}{\frac{x_{t-i}}{q}},
\end{equation}
$\{\theta_i~|~i~\in~1,\dots,q\}$ are parameters of the model and 
$\{\varepsilon_i~|~i~\in~1,\dots,q\}$ are the white noise error terms, which $
\varepsilon_i = x_{i} - \mu$.

\paragraph{Pros}

Using the MA method, events that influenced a value observed at time $t$ can 
only have influence on the most recent observations, i.e., the predictions do 
not follow short-term trends.
Moreover, this method can also be used to remove casual noise from the data, 
given that its predictions are less sensitive against outliers than those made 
using AR models, for example.

\paragraph{Cons}

In order to calculate the parameters of the model ($\theta_i$), it is not 
possible to use linear least squares, and iterative non-linear fitting 
procedures are required, which makes the MA method computationally more complex 
than the AR one.
Furthermore, similarly to AR, the accuracy of this method significantly 
decreases when predicting more values than its own order. 
This occurs because after the $q^{th}$ prediction, it does not have any actual 
observation to compare and the predictions tend to an average value that
usually does not match with the real observations.

\subsubsection{Exponential Smoothing (ES)}

The simplest version of the ES is also known as Exponentially Weighted Moving 
Average (EWMA) and the value predicted for the time $t + i$ can be calculated 
using only the most recent observation and the most recent forecast.
For instance, the value of $\hat{x}_{t}$ is the weighted average:

\begin{equation}
 \hat{x}_{t} = \alpha x_{t-1} + (1-\alpha) \hat{x}_{t-1}
\end{equation}

Guided by the value of $\alpha \in [0, 1]$ (also called \emph{smoothing 
constant}), the relevance of the old measurements undergo an exponential decay, 
which justifies its name.

Other formats of the ES are also widely used, adding up to two new parameters 
($\beta$ and $\gamma$) in order to better detect non-linear trends.
A common way to setup good values for $\alpha$, $\beta$ and $\gamma$ is by 
trying among $k$ possible values each (usually, $k=10$).
The choice is made according to the errors observed over the data already 
observed, e.g., calculating a prediction $\hat{x}_{t-1}$ and comparing it with 
the real observation ${x}_{t-1}$.

\paragraph{Pros}

The space and time complexities are smaller, when compared with the AR 
and MA methods, and the predictions incorporate better the trends in the last 
observed values.

\paragraph{Cons}

It has some of the same limitations observed in the MA methods, such as the low 
efficiency when predicting the data value even in short time intervals. 
Moreover, its confidence intervals increase exponentially.

\subsubsection{Autoregressive Integrated Moving Average (ARIMA)}

An ARIMA model defines a stationary process that is composed by the 
combination of an AR and a MA models. 
Values that will be observed in the future can be more accurately predicted 
if the calculation considers:
\begin{inparaenum}[(i)]
 \item the magnitude of the last observations and their trends (incorporated by 
the AR model); and 
 \item the impact of (unobserved) shocks that influenced their current state 
(incorporated by the MA model).
\end{inparaenum}
In case of having a non-stationary data, an initial differencing step 
(corresponding to the "integrated" part of the model) can be applied. 
Such a transformation can be represented by the equation:

\begin{equation}
 y_{t} = (1 - L)^d x_{t},
\end{equation}
where $d$ is the order of the integrated model and $L$ is the Lag operator, 
such that $L^k x_t = x_{t-k}$  for all $t > k$.

An $\text{ARIMA}(p,d,q)$ model contains an AR model with order $p$ 
and a MA model with order $q$ and a value observed at time $t$ can be 
represented as 

\begin{equation}
 y_{t} = c + \varepsilon_t + \sum_{i=1}^{p}{\alpha_{i} y_{t-i}} + 
\sum_{i=1}^{q}{\theta_i \varepsilon_{t-i}}.
\label{eq:arma}
\end{equation}

Note that this formula is used to predict the value of $y_t$, which is derived 
from $x_t$, if $d > 0$, or simply equal to $x_t$, if $d = 0$.
Moreover, as well as the AR, MA and ES methods, the parameters $p$, $q$ and $d$ 
are usually 
set among $k$ possible values each, using an information criterion measure 
(usually, $k \in [0,6]$).

\paragraph{Pros}

The ARIMA method has higher accuracy than the previous methods, given that the 
predictions consider new trends observed in the latest observations and 
converge much slower than the MA models to the average values.
Furthermore, this method has a particular characteristic which is the 
overlapping with other methods.
That is, some ARIMA models are equivalent to other methods, for example:
\begin{itemize}
 \item $\text{ARIMA}(0,1,0)$ is equivalent to the naive method that assumes the 
last observation will repeat in the future;
 \item $\text{ARIMA}(0,2,0)$ is equivalent to the naive method that assumes a 
linear increasing based on the last two observations;
 \item $\text{ARIMA}(0,1,1)$ is the simplest model of the ES (with 
only one parameter); and
 \item $\text{ARIMA}(0,2,2)$, $\text{ARIMA}(0,1,2)$ and $\text{ARIMA}(1,1,2)$ 
are equivalent to more complex ES models.
\end{itemize}
Hence, when choosing the ARIMA model that fits more properly to the data, some 
of the other methods are also implicitly considered.

\paragraph{Cons}

The time and space complexities are bounded by the worst complexity between the 
AR and the MA methods, which depends on the values of $p$ and $q$ (as shown in 
Table~\ref{table:time-series-methods}). 
Moreover, an extra step may be required to differentiate the data.

\subsection{Regression methods} 

Regression methods have a different approach than time series methods. 
Instead of relying only on past values to make predictions, they also predict 
measurements based on other measurement types. 
For instance, given a value observed by one sensor node, a regressive model can 
be used to predict which value would be observed by another sensor node.
In the following, we discuss three methods used in WSN environments: the linear 
regression, the kernel regression and the Principal Component Analysis (PCA).
Table~\ref{table:regression-methods} summarizes their characteristics and the
references that applied each of them.

\paragraph{Pros}

It is possible to combine different data types, e.g., to use temperature 
measurements to predict the relative humidity at the same time or some moments 
later.
In other words, regression methods consider that the environment is composed by 
more than a single type of data and that other factors may influence the 
studied process, which fits to the distributed architecture of WSNs.

\paragraph{Cons}

In comparison with time series methods, regression methods have higher space and 
time complexities that, given the constrained capacities of some sensor nodes, 
may limit their adoption in WSNs.

\subsubsection{Linear Regression}

Linear regressions are the simplest kind of regression.
They are used to characterize linear relations between the observed variables.
Using a linear regression, it is possible to use a measurement $x$ to predict 
the value of $y$ based on a linear function $y = \beta_0 + \beta_1 x$.
The coefficients $\beta_0$ and $\beta_1$ can be calculated using the least 
squares method (see~\cite{Diez2012} for other methods).


\paragraph{Pros}

It is possible to make three different types of 
predictions that are useful in WSN scenarios~(\cite{Deshpande2004}): 
\begin{inparaenum}[(i)]
 \item range based, i.e., to predict whether a measurement will be inside a 
range of values;
 \item value based, i.e., to calculate the probability that a measurement will 
be a certain value; and
 \item average aggregation, i.e., to predict the average of a set of unobserved 
measurements at time.
\end{inparaenum}

\paragraph{Cons}

Linear regressions assume normally distributed data, which may rarely happen 
when considering several sources of data and requires within-study correlation 
estimates.

\subsubsection{Kernel regression}

Kernel density estimation is a non-parametric model used to estimate the 
probability density function of the observed data, starting with no assumptions 
about the data distribution.
The goal of the kernel regression is to find the value of $ E[Y \vert X] = 
m(X)$ for an unknown function $m(\cdot)$.
To achieve that, a regression is made based on the given values of $X$, with 
the help of a kernel function that is responsible for quantifying the 
similarity between their data points.
Finally, a new probability density function is drawn based on the observed 
values and can be used to predict the value of $E[Y \vert X]$.

\paragraph{Pros}

It does not require any assumption about the data distribution and tend to have 
smaller errors when compared with the linear models. 

\paragraph{Cons}

To compensate the absence of assumptions, more data is required to find a proper 
approximation to the real distribution.
Hence, its computational complexity is much higher than of linear regressions, 
both in terms of space and time.

\subsubsection{Principal component analysis (PCA)}
\label{sec:pca}

The PCA method is used to reduce the dimensionality of data sets.
It uses orthogonal transformations to convert sets of observations of 
(possibly) correlated variables into sets of values of linearly uncorrelated 
variables, so-called \emph{principal components}.
After the conversion, only the principal components that retain most of the 
variation present in the data set are kept.
Thus, based on the information retained by these components, it is possible to 
predict the values of the original data set with a high degree of confidence.
For instance, it may be possible to predict the measurements of all the sensor 
nodes based on the measurements made by those that observe the highest 
variations in the environment.
The algorithm to calculate the principal components uses eigenvectors and 
a covariance matrix, and is thoroughly explained in~\cite{Jolliffe2002}.

\paragraph{Pros}

The PCA method provides a means to remove part of the data without losing the 
relevant information that it contains.

\paragraph{Cons}

The pre-processing phase involves the computation of the product of matrices 
based on large sets of data.
It means higher time and space complexities than the other options.  

\subsubsection{Gaussian Process regression}
\label{sec:gaussian-process}

In short, a Gaussian Process (GP) is a collection of random variables.
There, a finite set of such random variables has a joint multivariate Gaussian 
distribution, i.e., it is defined by their means and the covariance of 
the distributions.
Each random variable ($f(x)$) is indexed by $x$ and a covariance function 
that incorporates prior assumptions about the relation with the other
distributions. 
Because of that, no advanced knowledge about the data is required for making 
accurate predictions.
In~\cite{Rasmussen2006}, the GP regression method is explained in detail.

\paragraph{Pros}

The GP regression method produces probabilistic models that are composed by 
single values and associated confidence intervals, expanding their possible 
applications.
Furthermore, the GP regression is well-known because of their higher accuracy 
when compared to other regression methods, thanks to its fast fitting to the 
underlying (unknown) data distribution.


\paragraph{Cons}

The main drawback of the GP regression is the computation time required to 
make a prediction.
As it is shown in Table~\ref{table:regression-methods}, the computation 
has a cubic growth and new models cannot be generated online, which makes it an 
unfeasible option for larger datasets.
It is worth to mention that there are works focused on reducing its 
computational complexity in exchange for reducing the 
accuracy~(\cite{Saat2011}).

\subsection{Machine learning techniques}
\label{sec:machine-learning-techniques}

As shown in~\cite{Alsheikh2014}, machine learning techniques have been adopted 
in several solutions for WSNs at different levels, such as routing, medium 
access control and event detection.
However, from these solutions, only Artificial Neural Networks (ANNs) have been 
applied for reducing their number of transmissions.

\subsubsection{Artificial neural networks}
\label{sec:anns}

The methods described in the previous Sections are mainly considered 
\emph{traditional methods}, given that they are extensions of probabilistic 
approaches.
As described in~\cite{Jain1996,Haykin1999}, ANNs are based on a different 
paradigm and their general idea is to create an artificial version of the 
biological neurons.
That is, to simulate a network of components (so called \emph{neurons}) and 
predict a system's output, given a set of inputs.
To achieve that, an ANN must go through a \emph{learning phase}, i.e., to adapt 
its internal parameters and learn from available historical data.

\paragraph{Pros}

The main advantage of ANNs is that they are able to handle multiple data 
types and model regressions between several variables.
Additionally, ANNs have been observed to perform more accurately than the 
traditional methods in time series with discontinuities, which may happen in 
case of absence of parts of the data--very common in some WSNs. 
Finally,~\cite{Kang:1992:IUF:144978} found that ANNs often perform more 
accurately for long term predictions than for smaller intervals.

\paragraph{Cons}

ANNs are soft computing solutions that cannot be bounded by a computational 
time limit.
One of the reasons for the high computational costs is that the design process 
of an ANN involves selecting the number of hidden layers, adjusting the 
connection between each layer (the synapses), choosing the number of neurons in 
each layer, and setting the activation function, a learning algorithm and the 
number of training samples.

ANNs are powerful and can approximate other prediction methods and nonlinear 
models, given the best conditions, such as enough data.
However, the amount of data required to find a stable ANN is much higher than 
the others, because it has more parameters to estimate.
Moreover, there is no theoretical guarantee that they will perform well for 
out-of-sample forecasts, i.e., predictions after the learning phase.
In conclusion, it may be costly to find out which situations better fit to 
neural networks than to traditional models.

\section{Discussion}
\label{sec:discussion}

\input{table-refs-summary.aux}

In this Section, we answer some questions that may arise during the design of a 
strategy to adopt predictions to reduce the amount of transmissions in a WSN:
\begin{enumerate}
 \item How to improve a WSN using predictions?
 \item How to choose a prediction model?
 \item Why (not) make predictions in \BaseStation{s}?
 \item Why (not) make predictions in sensor nodes?
 \item Why (not) make predictions in \ClusterHead{s}?
\end{enumerate}

The discussion is supported by the statistical literature (partially 
represented by the prediction methods explained in 
Section~\ref{sec:prediction-methods}) and by the results presented in the works 
included in this survey.

\subsection{How to improve a WSN using predictions?}


Some characteristics of a WSN are less flexible than others.
For example, a WSN that was designed and deployed to track objects cannot be 
simply changed to monitor room temperature, because its sensors measure 
other parameters than temperature, relative humidity, solar radiation or any 
other value that could be correlated with the local temperature.
Therefore, the WSN type is a firm characteristic, because it would be highly 
costly to change sensors in nodes already deployed.

On the other hand, one may have some information about the data which is going 
to be monitored by a WSN at the moment of the deployment, but there are inherent 
costs to acquire real data from the environment and perform its analysis.
Such costs must be taken into account when calculating the improvements brought 
by the use of predictions, but they are not as impeding as changing the WSN 
type.

Therefore, to answer ``how to improve a WSN using predictions?'', we describe 
a bottom-up approach that goes from less flexible characteristics, such as the 
types of sensors available in a WSN, to the economically cheapest and the least 
time-consuming ones.

\subsubsection{WSN type}

There are three large classes of WSNs: one encompasses the networks used for 
event detection; another the WSNs for monitoring and reporting (query-based or 
continuously); and the networks used for object (and people) tracking.
Each of these classes has different requirements about timeliness, throughput
rate, computing power in sensor nodes and tolerance to packet losses.
Hence, for instance, it is unlikely to re-utilize in a fire-detection scenario 
a WSN that was originally deployed to monitor the temperature, due the economic 
and death risks that it would involve.
In other cases, it may nearly impossible to adapt one WSN to make another task, 
because their sensor nodes may not contain the proper hardware to measure a 
certain parameter, as explained before.

Therefore, the WSN type must be the first aspect to be taken into account when 
considering the use of predictions to reduce the number of transmissions in a 
WSN.
In the following, we list the requirements of each WSN type and give examples
of applications that can be used in each case.
As a reference, in Table~\ref{table:approaches}, we filled the second column 
with a letter \TableEvent{} to identify prediction-based solutions adopted 
in event detection applications, \TableQuery{} in query-based solutions, 
\TableContinuous{} for continuous monitoring and \TableObject{} for 
object tracking ones.

\paragraph{Event detection}

Such networks are very strict about the delays that the transmissions may 
suffer, because having an undesired delay when reporting an event may cause 
economic losses or put lives at risk, depending on the situation (e.g., in a 
disaster detection).
However, if all sensor nodes transmit at the same time, it may cause 
congestion. 
As a consequence, the number of packet collisions may increase and delay the 
delivery of relevant information. 
Therefore, these WSNs should avoid approaches that ignore the delay of the 
packets as a critical issue, due to the risks that it can bring to the 
environment surrounding the WSN.
It is also possible to adopt regression methods or machine learning techniques 
to predict events, which requires advanced knowledge about the domain under 
study and specific refinements according to the scenario.


\paragraph{Query-based monitoring}

Monitoring WSNs are used to measure environmental parameters, people's health,
machines and engineering structures with low computational power and energy
supplies.
Especially, query-based monitoring WSNs work in a pull-based fashion, i.e., 
users eventually demand information from the environment and the 
\ClusterHead{s} make an effort to answer most of the queries locally and 
transmit them as rarely as possible to sensor nodes. 
As in most cases user queries contain which information must be retrieved and
the error tolerated by the user, the \ClusterHead{s} may predict the sensors' 
measurements and avoid transmissions, whenever the confidence levels of the
predictions match to the users' expectations.

\paragraph{Continuous monitoring}

As in query-based WSNs, it is common to encounter temperature, relative 
humidity, light, solar radiation, wind speed and soil moisture sensors, among 
others, that can measure environmental parameters.
The main difference may be in the density of sensor nodes in a deployed
network.
If the WSN is set to continuously transmit the state of the environment, some
sensor nodes may run out of battery quicker and their data should be instantly 
replaced by the data collected in the same region.
Also, thanks to the data redundancy inherent to densely deployed WSNs, 
they usually have less restrictions about delays that may occur during the 
transmissions.
Hence, adaptive sampling mechanisms can use predictions to avoid unnecessary
transmissions and substitute the sensed data that has not been reported to 
\ClusterHead{s} in a DPS. 

\paragraph{Object tracking sensor networks (OTSNs)}

Some WSNs are responsible for tracking objects. 
Their tasks may vary from simply detecting the presence of a person in a 
region, to applications that track the (almost) exact position of an enemy in a 
battlefield, an animal in a farm or cars in a smart city.
These WSNs are less tolerant to delays in data delivery and require more 
detailed information from sensor nodes, such as a high-precision 
measurements~(\cite{StrategiesXu2004}).
Because of the importance of the detailed information, this kind of application 
usually requires sensors that consume more energy and are more expensive, such 
as cameras and microphones.

In order to keep the timeliness in the data delivery, it is important to avoid 
packet collisions and medium congestion.
Thus, transmissions are made by as less sensor nodes as possible at a 
time, differently from monitoring applications that keep collecting as more 
parameters as possible to build the most complete visualization of the 
environment in the \BaseStation{s}.
To do that, OTSNs adopt sensor nodes with higher computational resources, which 
are able to autonomously decide the best times to measure or 
transmit~(\cite{Bhatti2009}).
Hence, the computation in sensor nodes may be heavier, and a SPS with model 
generation in sensor nodes can help to meet their strict timeliness and reliable 
data delivery requirements.


\subsubsection{Energy resources}

Limited energy resources is one of the main constraints in WSNs. 
However, some of the works presented in this survey assume that some sensor 
nodes have larger energy resources than the others, which may occur thanks to 
more powerful (and therefore more expensive) batteries, energy harvesting or 
simply because they 
are plugged in.
In conclusion, \ClusterHead{s} may exploit the abundance of their energy 
resources to choose a prediction method that is more complex and probably more 
accurate than the others. 

As shown in Tables~\ref{table:time-series-methods} and 
\ref{table:regression-methods}, some methods have much higher runtime complexity 
than the others, e.g., the GP regression has a cubic growth while the ES grows 
linearly.
Therefore, if sensor nodes have scarce energy resources, it may be more 
rational to restrain the options to the least complex methods.
Finally, it is worth to mention that, besides the energy resources, the 
processing power of the sensor nodes plays an important role in the adoption of 
prediction methods.
We discuss such limitations in the following.

\subsubsection{Processor}

The computational power of the sensor nodes is determinant to decide which kind 
of operation they will perform.
Sensor nodes with capacity to compute complex operations can be set to generate 
a set of prediction models and choose the one that better fits to the current 
data, make predictions and compare with real measurements.
On the other hand, if the sensor nodes' processors cannot perform complex 
mathematical instructions, they may be limited to naive prediction methods, to 
predict values using models calculated by \ClusterHead{s} or to simply compare 
predictions computed by \ClusterHead{s} with real measurements.

Furthermore, the relation between the energy consumed when executing a machine 
instruction and the energy spent to make a radio transmission has been 
considered in few works, but has been shown as a relevant aspect by 
some authors.
That is, many authors assume that processing data is always less resource 
wasting than making a radio transmission.
However, in~\cite{Goel2001}, it has been shown that their approach is energy 
efficient only if a prediction can be computed using less than $15.000$ machine 
instructions in their environment.
Moreover, given that some sensors have much higher consumption than the others 
(especially in OTSNs, as explained in~\cite{Anastasi2009}), such a number can 
significantly vary from case to case.

In Table~\ref{table:approaches}, we used the third column to show the works 
that compute extra instructions in sensor nodes, for example, predictions 
and/or their models.
The fourth column shows checkmarks in the works that make extra transmissions, 
either because of the prediction models' parameters or the decision about 
adopting predictions, taken in runtime.
The fifth column shows the works that completely turn sensor nodes 
off (i.e., both the micro-controller unit and the embedded sensors), impacting 
the overall consumption provoked by sensing tasks and transmissions.


\subsubsection{Storage space}

WSNs are typically composed by cheap wireless sensor nodes and, in order to 
keep their costs low, their amount of available memory is extremely limited. 
In Tables~\ref{table:time-series-methods} and~\ref{table:regression-methods}, 
we show the space complexity to store a prediction model, which can affect the 
decision about adopting a certain prediction method or not, given the sensor 
nodes' limitations.
Alternatively, the storage space problem may be solved at some cost using 
external flash memory in sensor nodes (as done in~\cite{Li2009}), which 
significantly increases the number of possibilities for prediction methods and 
may improve the predictions' accuracy.

\subsubsection{Information about the location of the nodes}

Using current localization techniques for WSNs, it may be possible to 
calculate the sensor nodes' exact location on the Earth based on information 
retrieved from a Global Positioning System (GPS) device connected to the 
WSNs~(\cite{Mao2007}).
Alternative schemes can be used in cases when sensor nodes have no 
information about their absolute position on the surface of the Earth, but they 
are able to calculate their relative position inside the WSN on different 
levels of granularity~(\cite{Pereira2012}). 
For example, they may be able to calculate the distance to their closest 
neighbors or to assess how many hops they are away from the \BaseStation{}.

The level of details in the information about the sensor nodes' location may 
lead to decisions about which kind of prediction is going to be made.
For example, in case of having the relative position of the nodes, it may 
possible to predict an aggregated function (e.g., the average) of 
the measurements from sensor nodes placed in a certain region (as done 
in~\cite{Cheng2003}).
If no information about their location is available, predictions may be 
aggregated according to the sensor nodes' data similarity, which can be 
observed after a training phase (as done in~\cite{Carvalho2011Journal}). 
Yet, when no information about location is available, it is possible to opt for 
a solution in which no aggregation is done, where each sensor node is 
separately predicted by a different prediction model (as done 
in~\cite{Debono2008}).

As a reference, we used the sixth column in Table~\ref{table:approaches} 
to illustrate, in each work, how much information about the location of the 
sensor nodes was necessary.
One checkmark is shown if only the relative information was required, e.g., the 
number of hops from the \ClusterHead{} to the node or the WSN topology.
The presence of two checkmarks means that the exact position of the sensor 
nodes was used, e.g., their GPS position.
Finally, on the absence of a checkmark, no knowledge about their localization 
was used.

\subsubsection{Historical data availability}

Some prediction methods (e.g., ANNs) require large amounts of historical data 
to generate an accurate prediction model.
However, in some scenarios, the data that is going to be measured by the sensor 
nodes cannot be observed or studied before selecting the best prediction method.
This lack of information may reduce the options of possible prediction methods 
that can be successfully applied.
As a reference, we added in the last column of Table~\ref{table:approaches} the 
information about which works assumed a priori knowledge about the data that 
they will work on.

When no assumptions about the data can be made, their statistical 
characteristics are not available or the historical dataset is absent, a 
``learning phase'' may be required.
The ``learning phase'', similar to the ``initialization phase'' in the DPSs, is
a period during which the sensor nodes report all the data that they have 
generated to the \ClusterHead{s} (\cite{Santini2006}).
Adopting a periodic ``learning phase'' (e.g., once a day) can improve the 
predictions' accuracy or expose when some prediction models are not performing 
as accurate as before.
Finally, the inherent costs of this procedure must be included in the plan of 
selecting the most proper prediction method for a specific scenario.

\subsection{How to choose a prediction model?}

Considering the multiple options to make predictions in different device types, 
the choice of the prediction model may lead to a successful deployment or simply 
make it extremely inefficient.
Usually, a richer model can provide more accurate results, but it may require 
more communication among sensor nodes, larger memory buffers or more computing 
time. 
As recommended in~\cite{Lazaridis2003}, this choice must be done using 
experimentation, expert opinion or past experience to choose between competing 
models. It has been pointed out in~\cite{LeBorgne2007} that "an inadequate a 
priori choice of a prediction model can lead to poor prediction performances".

In the literature about statistical methods (\cite{Kong}), a common 
way to choose a model among a list of options is to reward their accuracy and, 
in change, penalize their selection according to the number of parameters used 
to compute the predictions. 

Therefore, the first step is to assess the predictions' accuracy. This can be 
done in several ways by using measures that are supposed to attest the quality 
of a prediction model in a certain use case.
Examples of such measures are the well-known Mean Square Error (MSE), the Root 
Mean Square Error (RMSE), the Mean Absolute Error (MAE), the Root Mean Square 
Error (RMSE), the Relative MAE (RelMAE), the Mean Absolute Percentage Error 
(MAPE), the symmetric Mean Absolute Percentage Error (sMAPE), among 
others (\cite{Hyndman2006}).
Once the accuracy has been measured, it is necessary a way to measure the 
relative quality of the model.
Methods such as Akaike Information Criterion (AIC--presented 
in~\cite{Akaike1974}) and the 
Bayes Information Criterion (BIC--presented in~\cite{Schwarz1978}) are some of 
the existing options.

In~\cite{Li2013}, the methods described above were used to choose the best ARIMA 
model to make the predictions in their tests.
However, as we described in Section~\ref{sec:wsn-environments}, WSNs have 
computational limitations that most networks do not have. If the chosen 
architecture requires that the prediction model must be fixed a priori (and 
cannot be adaptively chosen), the decision must be made based on a few aspects 
that have influence in the energy consumption of the sensor nodes.
For example, the number of messages generated by the scheme when the prediction 
fails and all the engineering concerns, such as the energy consumed to 
(re-)fit prediction models and, especially, to transmit their parameters.

To overcome the limitations of traditional methods,~\cite{Liu2005} created 
a way to select a prediction model that considers the percentage of transmitted 
measurements ($r$) and the user desired level of accuracy ($\alpha$).
Later,~\cite{Aderohunmu2013} designed an extended model for the Prediction Cost 
($\text{PC}$), which is more generic and also considers the computational costs 
of each algorithm in the sensor nodes with respect to 
their memory footprint ($Ec$):

\begin{equation}
\text{PC} = [~\alpha f(e)~+~(1 - \alpha) r~] \text{Ec,}
\end{equation}
where $e$ is the measure of the predictions' accuracy (e.g., MSE, RMSE, sMAPE) 
and $f(e)$ is the accuracy according to the chosen measure.





Furthermore, we observed that among aspects that may become extra costs as 
a consequence of the chosen prediction method, the most important are the 
computation time required to prepare the model, the required data assumptions 
and the computing power, including the extra memory required to make new 
predictions. 
As an example, some model parameters can be adapted on the fly by 
using adaptive filters (e.g., as done in~\cite{Santini2006}), and hence there is no
need to store large sets of past data. 
On the other hand, they may require a lot of extra computation and extra storage 
capacities from sensor nodes and \ClusterHead{s}.

In Tables~\ref{table:time-series-methods} and~\ref{table:regression-methods}, 
we organized the traditional methods described in 
Section~\ref{sec:prediction-methods} according to their type: time series and 
regression methods.
It is possible to observe that some methods have low space and time complexity 
(e.g., constant predictions and exponential smoothing).
However, some options (e.g., AR, MA and ARIMA) have similar preprocessing time 
complexity and the space required by their models, together with their 
accuracy, should influence the final decision about which one to adopt.
The ANN method has not been included in any table because there is no exact 
solution and no upper limit for its complexity.
Usually some heuristics are adopted, for example, the maximum number of runs, 
but its performance depends on each use case and may be adjusted according to 
the results obtained.

\subsection{Why (not) make predictions in \BaseStation{s}?}

As explained in Section~\ref{sec:sps}, \BaseStation{s} can be used to make 
predictions that are going to be used by the WSNs.
The first reason to make predictions in \BaseStation{s} is because they are 
supposed to be the most powerful device in WSNs (in terms of computing power 
and energy supply) and, therefore, they may be able to utilize more information 
about the environment, external changes and historical data.

\subsubsection{Pros}

Existing mechanisms show that it is not necessary to establish a communication 
between the \BaseStation{} and a sensor node in order to predict its values, 
i.e., it is possible to predict measurements that a sensor node is going to 
make based on its neighbors' measurements. 
As a consequence of this, a sensor node can have its MCU completely turned off 
for a while, which may be an important step on the direction of improving the 
overall WSN lifetime.
Furthermore, since \BaseStation{s} have access to information retrieved from 
several locations, they are able to see the broad picture of the environment, 
better understand how it is evolving and infer predictive models at a larger 
time-space scale, accounting for possibly existing cyclic behavior, global 
trends or other aspects not discernible from the sensors' limited (time and 
space-wise) perspective.

\subsubsection{Cons}

On the other hand, predicting future measurements in \BaseStation{s} has also 
disadvantages. 
For example, usually, values available in \BaseStation{s} are actually based on 
the estimations from old measurements and not on the most recent values.
Hence, bad assumptions about the data distribution may lead to inaccurate 
predictions that significantly worsen the quality of information delivered 
by the \BaseStation{} to the WSN owner.

There is another disadvantage about the quality of the delivered information. 
Let us suppose that at time $t$ the \BaseStation{} predicts the measurement 
that a sensor node will make at time $t+1$. 
If the difference between $t+1$ and $t$ is smaller than the delay to retrieve 
measurements from sensor nodes, the system will fail to keep the quality of 
its information when a prediction is not accurate, because it will not have 
time enough to request the actual measurements made in that instant.
For example, a system predicts the measurement that will be made in one minute; 
if its confidence interval does not match to the user requirements, it will 
have to request the real measurement from the sensor nodes;
however, if the total time to request and receive the measurement is greater 
than one minute, the \BaseStation{} will fail to attend the confidence level 
defined by the user.


\subsection{Why (not) make predictions in sensor nodes?}

As we showed in Sections~\ref{sec:sps} and~\ref{sec:dps}, it is possible to use 
the sensor nodes not only to measure information from the external world, but 
also to make predictions and reduce the number of transmissions in a WSN.

\subsubsection{Pros}

SPSs can avoid unnecessary measurements in sensor nodes that track objects and 
extend their lifetime, because they are usually equipped with more powerful 
sensors, such as cameras, microphones and radio-frequency identification, that 
require more energy to make a measurement than to process a few machine 
instructions~(\cite{Raghunathan2006}).

Regarding the DPSs, the main advantage of exploiting the computing power of the 
sensor nodes is that since they produce raw data series, the prediction quality 
can be tested without implying on high communication costs, because the sensor 
nodes are able to check their accuracy locally.
Moreover, a recent study has shown that it is possible to successfully 
substitute real measurements by predictions, reducing the number of 
transmissions without affecting the quality of the measurements provided by the 
WSN~(\cite{Dias2016}).

\subsubsection{Cons}

SPSs do not fit to sensors used to monitor environmental parameters, such as 
temperature, relative humidity and solar radiation, because they might spend 
more energy to predict than to sample the environment.
Moreover, since the computing power of these sensor nodes is usually limited, 
they may not incorporate information about distant past because of the lack of 
memory space. 
As a consequence of the limited computing power and the low amount of 
information available, complex prediction models may become unfeasible.

Besides that, the absence of updates arriving at the \BaseStation{} may 
incorrectly imply that predictions are accurate, which requires extra 
transmissions (e.g., beacons) to assure their activity status. 
Therefore, making predictions in sensor nodes requires a reliable transmission 
scheme in order to make it possible for \BaseStation{s} to provide accurate 
information.


\subsection{Why (not) make predictions in \ClusterHead{s}?}

As described in Section~\ref{sec:wsn-environments}, 
\ClusterHead{s} can be viewed as local \BaseStation{s} placed closer to 
sensor nodes, because they are responsible for the communication between the 
sensor nodes and the \BaseStation{} in a WSN.
The extra responsibility usually results on higher energy consumption in 
\ClusterHead{s} than in ordinary sensor nodes, provoked by the higher number 
of transmissions.


If \ClusterHead{s} have higher resource availability, it is possible to use more 
sophisticated prediction methods and exploit the same advantages of making 
predictions in \BaseStation{s}.
In such cases, making predictions in \ClusterHead{s} reduces the number of 
transmissions in WSNs, \textbf{improve} the communication in the clusters and 
\textbf{extend} the sensor nodes' lifetime.
In comparison with making predictions in \BaseStation{s}, now the time spent 
to retrieve measurements from sensor nodes is shorter, thanks to the lower 
number of hops between \ClusterHead{s} and sensor nodes.

Alternatively, some clustering methods are able to periodically elect the 
\ClusterHead{}~(\cite{Younis2006}). 
In such cases, when a sensor node runs out of battery, it is expected a 
decrease in the accuracy of the measurements made by its cluster. 
Thus, using an accurate prediction model to reduce the number of 
transmissions may have a larger impact in the quality of the information 
provided by the WSN.
Furthermore, because of the extra responsibilities and the higher number of 
transmissions, if ordinary sensor nodes act as \ClusterHead{s}, their own 
lifetime will be sharply reduced.
Thus, differently from the previous case, adopting a data reduction scheme may 
be necessary to \textbf{keep} the WSN alive for a reasonable time.


Note that making predictions in \ClusterHead{s} does not preclude predictions in 
\BaseStation{s} nor in sensor nodes. 
It is possible to make predictions to reduce the transmissions from 
\ClusterHead{s} to \BaseStation{s} 
and to reduce the transmissions between sensor nodes and \ClusterHead{s}, as 
done in~\cite{Goel2001,Wu2016}. 

\section{Open Issues \& Future challenges}
\label{sec:future-challenges}


The first researches about prediction-based data reduction in WSNs focused on 
the most economic ways to process the sensed data and avoid unnecessary 
transmissions.
From these works, two main architectures emerged:\begin{inparaenum}[(i)]
\item in SPSs, either the \ClusterHead{s} or the sensor nodes make all 
predictions, relying on the confidence assessed by the chosen prediction 
method; and
\item in DPSs, sensor nodes and \ClusterHead{s} make predictions 
simultaneously and exploit the sensor nodes' proximity to the origin of the 
data to avoid unnecessary transmissions.
\end{inparaenum}

We observed that most of the recent works tended to use DPSs to exploit the 
best characteristics of each WSN component, i.e., the extra computing 
power of \BaseStation{s} and \ClusterHead{s}, and the sensor nodes' proximity 
to the sources of data.
Therefore, the newest contributions focused on finding the most proper
prediction methods that:\begin{inparaenum}[(i)]
\item fit to limitations imposed by WSNs; and
\item have high precision in several scenarios, such as indoor and outdoor 
environment monitoring, precision agriculture and structural health monitoring, 
etc..
\end{inparaenum}
Prediction methods that predict accurately in several scenario are preferred, 
because they have higher chances of keeping the high accuracy in new 
deployments.

The work described in~\cite{Dias2015} shows the potential of data reduction in 
an average scenario with a DPS: high accuracy forecasts can reduce up to $30\%$ 
the number of transmissions.
The described model can be used as a baseline for future works that focus on 
the data plane to reduce the number of transmissions in sensor networks.

The biggest challenges for the future works involve incorporating 
characteristics from the statistical theory, considering heterogeneous computing 
power capabilities and managing the large-scale use of predictions. 
These topics are discussed in the following.

\subsection{Statistical theory}

In general, current works that use predictions do not refer to any mechanism 
for data analysis. 
Hence, statistics considered for data analysis are built before choosing a 
prediction method and computing prediction models.
That is, the environment is supposed to evolve and change in time 
(\cite{Aderohunmu2013Impl,Aderohunmu2013}), but this is not considered in 
many cases, such as in~\cite{Guestrin2004,Li2013}.

Most of the surveyed works use predictions models in their solutions with very 
low or absolutely no mathematical basis, i.e., authors usually ignore the 
existence of related works in statistics when deciding which prediction method 
can be the best one for their scenario, which decreases the reliability of 
their mechanisms.
For example, most of the works 
(\cite{Yann-Ael2005,Tulone2006,Deshpande2004,Chu2006,Guestrin2004,Debono2008,Jiang2011,Min2010,Askari2011,Stojkoska2011,Carvalho2011Journal,Aderohunmu2013Impl,Yin2015,Raza2015,Wu2016}) are based on the dataset from the 
experiments described in~\cite{IntelLabData:2004:Misc}. 
However, each work takes its own decision about which prediction method to 
use, i.e., none of them incorporate tools to properly analyze the data and find 
out its characteristics before choosing the prediction method that best fits 
to their requirements.
Future approaches may consider the existence of the other prediction methods 
and the possibility of choosing different methods according to the current state 
of the environment and its evolution.

Some authors support their assumptions on the statistical theory, but do not 
consider other details inherent to WSNs.
For instance, a mechanism to select a prediction model that considers some 
characteristics of the WSN environments has been presented 
in~\cite{Aderohunmu2013Impl,Aderohunmu2013}, but the authors did not 
incorporate the elevated costs to (re)transmit prediction models and their 
parameters.
In conclusion, one challenge for future works is to develop a mechanism to 
evaluate the efficiency of the use of predictions for data reduction. 
Differently from the current solutions, the new mechanism should be based not 
only on the communication costs in the evaluation process (as done in 
\cite{Jiang2011}), but also on 
\begin{inparaenum}[(i)] 
	\item the prediction model chosen;
	\item where the predictions will be computed; and 
	\item the processing costs implied in their computation.
\end{inparaenum}

Adding statistical tools into a mechanism that handles WSNs may be challenging, 
because it involves two distinct areas of knowledge (statistics and computer 
networks) and a new solution would require a high affinity between their 
advantages and disadvantages.
On the other hand, a mechanism will be significantly more reliable if its 
actions are based on a statistical theory study of the measured data.
Among other benefits, it will be possible to assess its potential improvements 
in the medium congestion and WSNs' lifetime, as well as lower bounds for the 
quality of the information that it would produce.

\subsection{Heterogeneous computing power capabilities}

As we described in Section~\ref{sec:wsn-environments}, sensor nodes, 
\ClusterHead{s} and \BaseStation{s} are expected to have different resources. 
This may happen not only because of their size, but also due to their roles in 
the system.
For example, it is expected that both computational and energy resources of 
\BaseStation{s} are several orders of magnitude larger than those of the sensor 
nodes. 
In other cases, such as in the Internet of Things (IoT), heterogeneous networks 
may be composed by sensor nodes with distinct memory and computing capacities.

It is an open challenge to exploit the strengths of the different devices in 
different ways. 
One possibility is to have different prediction methods running in the same WSN. 
For example, naive predictions running in sensor nodes with less resource 
availability, and more complex predictions (e.g., ANNs) in \ClusterHead{s}.

Another possibility is to explore the asymmetric characteristic from 
some prediction methods that use more computationally-intensive algorithms 
to compute models than to make predictions.
The ARIMA and the Kernel regression methods are examples of this asymmetry, as 
shown in Tables~\ref{table:time-series-methods} 
and~\ref{table:regression-methods}.
The work done in~\cite{Li2009} does a similar work based on the ARIMA method 
using \ClusterHead{s} to build and transmit prediction models to sensor nodes, 
which are responsible for making predictions.
This overcomes the current works that take binary decisions and either 
adopt more complex prediction mechanisms in SPSs, or build simpler prediction 
models when using dual prediction schemes. 


\subsection{Long term predictions}

Existing works are mainly focus on predicting measurements that are going to 
be done in the short term, for example, in the next $5$ minutes.
The constrained time interval is chosen according to the limitation of the 
prediction methods (such as AR, MA and ARIMA), which needs to be often updated 
in order to produce accurate predictions.
In DPSs, inaccurate predictions make WSNs consume much more energy to recompute 
prediction model parameters and transmit the updates through the network.

Predicting longer time-intervals (for example, one hour) may provide a 
perspective about when the accuracy of the short term predictions will decrease.
That is, based on the extra computational power of the \BaseStation{}, it may 
be possible to anticipate whether the information produced by the WSN will not 
meet the minimum quality requirements in the future.
As a response, prediction models may be updated or new prediction schemes can 
be adopted before a decrease in quality is actually observed.

We expect that long term predictions may be feasible by means of using external 
information, which can be either from other WSNs or from third-part 
sources~(\cite{Oechsner2014}).
This is clearly not a trivial question, since decisions taken at this level 
may involve extra transmissions and processing costs to change the sensor 
nodes' operation.
As a trade-off, it may bring benefits that are not in question in the actual 
state of the art, in terms of energy savings, medium access and quality of 
information.
For instance, leading to a new set of mechanisms that do not focus only on 
extending the WSNs' lifetime, but also on providing more information to
users, which has not been considered as a possibility within the use of 
predictions, so far.



\section{Conclusion}
\label{sec:conclusion}

The study of the state-of-the-art reveals that there is a good reason to invest 
in making predictions in WSN environments that goes beyond adapting the 
systems' operation to save energy resources and extending the WSNs' lifetime.
The future of the IoT depends on the scalability of sensor networks and their 
capacity to autonomously manage their access to the wireless medium.

With this work, we aim to reduce the gap between the fields of statistics and 
WSN management, and expect that future works in the WSN and IoT fields will be 
improved by better applying the advantages of the predictions against the 
limitations of the WSNs.
To achieve this goal, we surveyed the existing approaches that use predictions 
to reduce the number of transmissions in WSNs, and explained the prediction 
techniques that are currently being considered as options in WSNs. 

For the complete analysis, we categorized current works according to their 
architecture, highlighting the challenges to design new mechanisms.
For instance, any change in the WSNs' operation must handle two issues: it must 
detect sensor malfunctioning or changes in the reading dependencies among their 
measurements; and distribute energy consumption among the sensor nodes, which 
depends on their own predictability and can be unfeasible in some cases.
Moreover, we have also shown workarounds for the sensor nodes hardware 
limitations, such as extending the sensor nodes' memory capacity and avoiding,  
in sensor nodes, prediction methods that require complex machine instructions.

Finally, we observed that it is feasible to adopt predictions to reduce the 
number of transmissions in WSNs.
However, it depends not only on the predictions' accuracy, but also on the WSN 
goals, on the sensed phenomena, on the user requirements and on the architecture 
adopted to make the predictions.

\section*{Acknowledgment}

This work has been partially supported by the Spanish Government through the 
project TEC2012-32354 (Plan Nacional I+D), by the Catalan Government 
through the project SGR-2014-1173 and by the European Union through the 
project FP7-SME-2013-605073-ENTOMATIC.
We wish to thank the reviewers for their insightful comments and 
suggestions that allowed us to improve this paper.

\bibliographystyle{ACM-Reference-Format-Journals}
\bibliography{bibliography}

\end{document}